\def\verPreprint{1}
\def\verPAPER{2}
\def\ver{1}

\ifx\ver\verPreprint
\documentclass[a4paper,english,11pt]{article}
\usepackage[bindingoffset=0.5cm,left=2.5cm,right=2.5cm,top=2.5cm,bottom=2.5cm,footskip=1.0cm]{geometry}
\usepackage{lineno,amsmath,bm,multirow,amssymb,authblk,graphicx,newclude,xspace,hyperref,rotating}
\fi
\ifx\ver\verPAPER
\documentclass[1p]{elsarticle}
\usepackage{lineno,hyperref,amsmath,hepnames,bm,multirow,amssymb,xspace,rotating}
\fi

\modulolinenumbers[5]

\ifx\ver\verPreprint
\newcommand{\PW}{\ensuremath{\textrm{W}}\xspace}

\newcommand{\PGg}{\ensuremath{\gamma}\xspace}
\newcommand{\PHiggs}{\ensuremath{\textrm{H}}\xspace}
\newcommand{\Pp}{\ensuremath{\textrm{p}}\xspace}

\newcommand{\Pgt}{\ensuremath{\tau}\xspace}

\newcommand{\Ptop}{\ensuremath{\textrm{t}}\xspace}
\newcommand{\APtop}{\ensuremath{\bar{\textrm{t}}}\xspace}
\fi

\newcommand{\Pggx}{\ensuremath{\PGg^{*}}\xspace}

\newcommand{\pT}{\ensuremath{p_{\textrm{T}}}\xspace}
\newcommand{\pThat}{\ensuremath{\hat{p}_{\textrm{T}}}\xspace}
\newcommand{\kt}{\ensuremath{k_{\textrm{t}}}\xspace}

\newcommand{\HT}{\ensuremath{H_{\mathrm{T}}}\xspace}
\newcommand{\GeV}{\ensuremath{\textrm{GeV}}\xspace}
\newcommand{\TeV}{\ensuremath{\textrm{TeV}}\xspace}
\newcommand{\data}{\ensuremath{\textrm{data}}\xspace}

\newcommand{\incl}{\ensuremath{\textrm{incl}}\xspace}
\newcommand{\jet}{\ensuremath{\textrm{jet}}\xspace}
\newcommand{\pileup}{\ensuremath{\textrm{PU}}\xspace}
\newcommand{\Poisson}{\ensuremath{\textrm{Poisson}}\xspace}
\newcommand{\Nbar}{\ensuremath{\bar{N}}\xspace}
\newcommand{\wbar}{\ensuremath{\bar{w}}\xspace}

\newcommand{\MGvATNLO}{\textsc{MadGraph5}\_aMC@NLO\xspace}
\newcommand{\PYTHIA}{\textsc{Pythia}\xspace}
\newcommand{\HERWIG}{\textsc{Herwig}\xspace}

\newcommand{\POWHEG}{\textsc{Powheg}\xspace}

\newcommand{\SHERPA}{\textsc{Sherpa}\xspace}

\newcommand{\ie}{i.e.\xspace}
\newcommand{\eg}{e.g.\xspace}
\newcommand{\fbinv}{\ensuremath{\textrm{~fb}^{-1}}\xspace}

\usepackage{array}
\newcolumntype{C}[1]{>{\centering\arraybackslash}p{#1}}
\begin{document}

\ifx\ver\verPAPER
\begin{frontmatter}
\fi

\title{Stitching Monte Carlo samples}

%% Group authors per affiliation:

\ifx\ver\verPreprint
\author[1]{Karl Ehat\"aht}
\author[1]{Christian Veelken}
\affil[1]{National Institute for Chemical Physics and Biophysics, 10143 Tallinn, Estonia}
\fi
\ifx\ver\verPAPER
\author[tallinn]{Karl Ehat\"aht}
\ead{karl.ehataht@cern.ch}
\author[tallinn]{Christian Veelken}
\ead{christian.veelken@cern.ch}
\address[tallinn]{National Institute for Chemical Physics and Biophysics, 10143 Tallinn, Estonia}
\fi

\ifx\ver\verPreprint
\maketitle
\fi

\begin{abstract}
Monte Carlo (MC) simulations are extensively used for various purposes in modern high-energy physics (HEP) experiments.
Precision measurements of established Standard Model processes or searches for new physics often require the collection of vast amounts of data.
It is often difficult to produce MC samples containing an adequate number of events to allow for a meaningful comparison with the data, 
as substantial computing resources are required to produce and store such samples.
One solution often employed when producing MC samples for HEP experiments 
is to partition the phase space of particle interactions into multiple regions 
and produce the MC samples separately for each region.
This approach allows to adapt the size of the MC samples to the needs of physics analyses that are performed in these regions.
In this paper we present a procedure for combining MC samples that overlap in phase space.
The procedure is based on applying suitably chosen weights to the simulated events.
We refer to the procedure as ``stitching''.
The paper includes different examples for applying the procedure to simulated proton-proton collisions at the CERN Large Hadron Collider.
\end{abstract}

\ifx\ver\verPAPER
\end{frontmatter}
\fi

\clearpage

%\linenumbers

%\begingroup
%\let\clearpage\relax
\section{Introduction}
\label{sec:introduction}

Monte Carlo (MC) simulations~\cite{Kroese2014WhyTM,dunn2011exploring} are used for a plethora of different purposes in contemporary high-energy physics (HEP) experiments.
Applications for experiments currently in operation include detector calibration; optimization of analysis techniques, including the training of machine learning algorithms;
the modelling of backgrounds, as well as the modelling of signal acceptance and efficiency.
Besides, MC simulations are extensively used for detector development and for estimating the physics reach of experiments that are presently in construction or planned in the future.

The production of MC samples containing a sufficient number of events often poses a material challenge 
in terms of the computing resources required to produce and store such samples~\cite{HSFPhysicsEventGeneratorWG:2020gxw}.
This is especially true for experiments at the CERN Large Hadron Collider (LHC)~\cite{Bruning:2004ej,Buning:2004wk,Benedikt:2004wm},
firstly due to the large cross section for proton-proton ($\Pp\Pp$) scattering and secondly due to the large luminosity delivered by the LHC.

The number of $\Pp\Pp$ scattering interactions, $N_{\data}$, that occur within a given interval of time 
is given by the product of the $\Pp\Pp$ scattering cross section, $\sigma$, and of the integrated luminosity, $L$, that the LHC has delivered during this time:
$N_{\data} = \sigma \, L$.
We refer to the ensemble of $\Pp\Pp$ scattering interactions that occur within the same crossing of the proton bunches as an ``event''.
The interaction with the highest momentum exchange between the protons is referred to as the ``hard-scatter'' interaction,
and the remaining interactions are referred to as ``pileup''.
The inelastic $\Pp\Pp$ scattering cross section at the center-of-mass energy of $\sqrt{s}=13$~\TeV, the energy achieved during the recently completed Run $2$ of the LHC (in the period $2015$-$2018$),
amounts to $\approx 75$~mb~\cite{Aaboud:2016mmw,Sirunyan:2018nqx}.
The $\Pp\Pp$ scattering data recorded by the ATLAS and CMS experiments during LHC Run $2$ 
amounts to an integrated luminosity of $\approx 140\fbinv$ per experiment~\cite{ATLAS-CONF-2019-021,LUM-17-001,LUM-17-004,LUM-18-002}.
Thus, $N_{\data} \approx 10^{16}$ inelastic $\Pp\Pp$ scattering interactions occurred in each of the two experiments during this time.
Ideally, one would want the number of simulated events to be higher than the number of events in the data,
such that the statistical uncertainties on the MC simulation are small compared to the statistical uncertainties on the data.
The production of such large MC samples is clearly prohibitive, however.

Even if one restricts the production of MC samples to processes with a cross section that is significantly smaller than the inelastic $\Pp\Pp$ scattering cross section,
such as Drell-Yan (DY) production, the production of $\PW$ bosons ($\PW$+jets), and the production of top quark pairs ($\Ptop\APtop$+jets),
the production of MC samples containing a sufficient number of events to allow for a meaningful comparison with the data represents a formidable challenge.
The DY, $\PW$+jets, and $\Ptop\APtop$+jets production processes are used for detector calibration and Standard Model (SM) precision measurements.
They also constitute relevant backgrounds to searches for physics beyond the SM.
Their cross sections amount to $6.08$~nb for DY production, $61.5$~nb for $\PW$+jets production,
and $832$~pb for $\Ptop\APtop$+jets production~\cite{Melnikov:2006kv,Li:2012wna,Czakon:2011xx}~\footnote{
  The quoted cross sections refer to, respectively, DY production of lepton (electron, muon, and $\Pgt$) pairs of mass $> 50$~\GeV,
  $\PW$+jets production with subsequent leptonic decay of the $\PW$ boson,
  and to the pair production of top quarks of mass $172.5$~\GeV.
}.
The ATLAS and CMS experiments would each need to produce MC samples containing $840$ million DY, $8.61$ billion $\PW$+jets, and $116$ million $\Ptop\APtop$+jets events
in order to reduce the statistical uncertainties on the MC simulation to the same level as the uncertainties on the LHC Run $2$ data.

In order to mitigate the effect of limited computing resources, both experiments employ sophisticated strategies for the production of MC samples.
A common feature of these strategies is to vary the expenditure of computing resources across phase space (PS),
depending on the needs of physics analyses.
When searching for new physics, for example, it is important to produce sufficiently many events in the tails of distributions,
as otherwise potential signals may be obscured by the statistical uncertainties on the SM background.

Different mechanisms for adapting the expenditure of computing resources to the needs of physics analyses have been proposed in the literature.
Modern MC programs (``generators'') such as \POWHEG~\cite{POWHEG1,POWHEG2,POWHEG3}, \MGvATNLO~\cite{MGvATNLO}, \SHERPA~\cite{SHERPA}, \PYTHIA~\cite{PYTHIA}, and \HERWIG~\cite{HERWIG}
provide functionality that allows to adjust the number of events sampled in different regions of PS through user-defined weighting functions.
This approach has been used in Ref.~\cite{ATLAS:2021yza}.
An alternative approach is to partition the PS into distinct regions (``slices'') and to produce separate independent MC samples covering each slice.
Following Ref.~\cite{HSFPhysicsEventGeneratorWG:2020gxw}, we refer to the first approach as ``biasing'' and to the second one as ``slicing''.

In this paper, we focus on the case that MC samples have already been produced and present a method that makes optimal use of these samples,
where ``optimal'' refers to yielding the lowest statistical uncertainty on the signal or background estimate that is obtained from these samples.
The samples in general overlap in PS.
For example, one set of MC samples may partition the PS based on the number of jets, 
whereas another set of samples may partition the PS based on $\HT$, the scalar sum in $\pT$ of these jets.
Our method is general enough to handle arbitrary overlaps between these samples.
The overlap is accounted for by applying appropriately chosen weights to the simulated events.
We refer to the procedure as ``stitching''.
One useful feature of the stitching method is that it allows to increase the number of simulated events incrementally in certain regions of PS
in case these regions are not yet sufficiently populated by the existing MC samples.

In the following, we will assume that all MC samples that are subject to the stitching procedure 
have been produced with the same version of the MC program and consistent (\ie identical) settings 
for parton distribution functions, scale choices, parton-shower and underlying-event tunes, etc.
In case a given set of MC samples was produced with inconsistent settings,
the effect of the inconsistencies either need to be small (compared to \eg the systematic uncertainties) or the events need to be reweighted
to make all MC samples consistent prior to applying the stitching procedure.

Variants of the stitching procedure described in the first part of this manuscript have been used by the ATLAS and CMS experiments since LHC Run $1$,
but, to the best of our knowledge, have not been described in detail in a public document yet.
The formalism for the computation of stitching weights is detailed in Section~\ref{sec:stitching_weights}.
Concrete examples for using the formalism in physics analyses are given in Sections~\ref{sec:WJets_vs_Njet} and~\ref{sec:WJets_vs_Njet_and_HT}.
The examples characterize the use of the stitching procedure by the CMS experiment during LHC Runs $1$ and $2$.
They are chosen with the intention to provide a reference.
In Section~\ref{sec:examples_trigger_rate} we extend the stitching procedure to the case of estimating trigger rates at the High-Luminosity LHC (HL-LHC)~\cite{TDR_Phase2_LHC},
scheduled to start operation in $2027$.
The distinguishing feature between the applications of the stitching procedure described in Sections~\ref{sec:examples_background_yield} and~\ref{sec:examples_trigger_rate}
is that in the former (but not in the latter) the cross section of the process that is modeled by the MC simulation
is orders of magnitude smaller compared to the inelastic $\Pp\Pp$ scattering cross section.
In the former case one can make the simplifying assumption that the process of interest (the process modeled by the MC simulation) 
solely occurs in the hard-scatter interaction and not in pileup interactions.
For the purpose of estimating trigger rates, a relevant use case is that the hard-scatter interaction as well as the pileup interactions are inelastic $\Pp\Pp$ scattering interactions,
and the hard-scatter interaction is in fact indistinguishable from the pileup.
As described in detail in Section~\ref{sec:examples_trigger_rate},
we account for this indistinguishability by making suitable modifications to the formalism for the computation of stitching weights.
The modified stitching procedure detailed in Section~\ref{sec:examples_trigger_rate} has been used to estimate trigger rates 
for the HL-LHC upgrade technical design report of the CMS experiment~\cite{TDR-21-001}.
We conclude the paper with a summary in Section~\ref{sec:summary}.

\section{Computation of stitching weights}
\label{sec:stitching_weights}

As explained in the introduction,
contemporary HEP experiments often employ MC production schemes
that first partition the PS into multiple regions and then produce separate MC samples covering each region.
We use the term ``MC production scheme'' to refer to the strategy for choosing which MC samples to produce and how to produce these samples 
(which MC generator programs to use, how to partition the PS into regions, which settings to use when executing the MC generator programs, etc)
and the term ``MC sample'' to refer to the set of all output files produced by one execution of a MC generator program.
When using these MC samples in physics analyses,
the overlap of the samples in PS needs to be accounted for by applying weights to the simulated events.
The weights need to be chosen such that the weighted sum of simulated events in each region $i$ of PS 
matches the SM prediction in that region:
\begin{equation}
\sum_{j} \, P_{j}^{i} \, s_{j}^{i} \, \sum_{k=1}^{N_{j}} \, w_{j}^{k} = L \, \sigma^{i} \, ,
\label{eq:one}
\end{equation}
where the symbol $L$ corresponds to the integrated luminosity of the analyzed dataset
and $\sigma^{i}$ denotes the fiducial cross section for the process under study in the PS region $i$.
The first (second) sum on the left-hand side extends over the MC samples $j$ 
(over the events $k$ in the $j$-th MC sample, where $N_{j}$ denotes the total number of simulated events in the sample $j$).
The symbol $w_{j}^{k}$ denotes the weight assigned to event $k$ by the MC generator program,
while $s_{j}^{i}$ denotes the ``stitching'' weight that is applied to events from the sample $j$ falling into the PS region $i$.
The symbol $P_{j}^{i}$ corresponds to the probability for an event in MC sample $j$ to fall into PS region $i$.
Eq.~(\ref{eq:one}) holds separately for each signal or background process under study.

One can show that the statistical uncertainty on the signal or background estimate
gets reduced when all simulated events that fall into PS region $i$ have the same weight,
regardless of which MC sample $j$ contains the event.
We hence choose the stitching weight to depend only on the PS region $i$ (and not on the MC sample $j$)
and refer to these weights using the symbol $s^{i}$ from now on.

We define the symbol $P_{\incl}^{i}$ as the ratio of the fiducial cross section $\sigma^{i}$ to the ``inclusive'' cross section $\sigma_{\incl}$,
which refers to the whole PS:
\begin{equation*}
P_{\incl}^{i} = \frac{\sigma^{i}}{\sigma_{\incl}} \quad \Longleftrightarrow \quad \sigma^{i} = \sigma_{\incl} \, P_{\incl}^{i} \, .
\label{eq:two}
\end{equation*}
Upon inserting this relation into Eq.~(\ref{eq:one}) and solving for the weight $s^{i}$, we obtain:
\begin{equation}
s^{i} = \frac{L \, \sigma_{\incl} \, P_{\incl}^{i}}{P_{j}^{i} \, \sum_{k=1}^{N_{j}} \, w_{j}^{k}} \, .
\label{eq:master}
\end{equation}

A special case, which is frequently encountered in practice,
is that one MC sample covers the whole PS,
while additional MC samples are used to reduce the statistical uncertainties in the tails of distributions.
We refer to the MC sample that covers the whole PS as the ``inclusive'' sample and the corresponding PS as the ``inclusive'' PS.
In this case, Eq.~(\ref{eq:master}) can be rewritten in the form:
\begin{equation}
s^{i} = \frac{L \, \sigma_{\incl}}{\sum_{k=1}^{N_{\incl}} \, w_{\incl}^{k}} \, \frac{P_{\incl}^{i} \, \sum_{k=1}^{N_{\incl}} \, w_{\incl}^{k}}{P_{\incl}^{i} \, \sum_{k=1}^{N_{\incl}} \, w_{\incl}^{k} + \sum_{j} \, P_{j}^{i} \, \sum_{k=1}^{N_{j}} \, w_{j}^{k}} \, ,
\label{eq:weight_incl}
\end{equation}
where $w_{\incl}^{k}$ refers to the weights assigned to events in the inclusive sample by the MC generator program and $N_{\incl}$ denotes the total number of events in the inclusive sample.
The sum over $j$ in Eq.~(\ref{eq:weight_incl}) extends over the additional MC samples, which each cover a different region in PS.
We will refer to these samples as the ``exclusive'' samples.
We assume that the weights $w_{\incl}^{k}$ and $w_{j}^{k}$ are normalized such that the average of these weights,
$\wbar_{\incl} = \frac{1}{N_{\incl}} \, \sum_{k=1}^{N_{\incl}} \, w_{\incl}^{k}$ and $\wbar_{j} = \frac{1}{N_{j}} \, \sum_{k=1}^{N_{j}} \, w_{j}^{k}$,
equals unity for the inclusive sample and for each exclusive sample $j$~\footnote{
  If this is not the case for a given set of MC samples,
  it can be achieved by a simple multiplication of the weights $w_{\incl}^{k}$ and $w_{j}^{k}$ by the factors $1/\wbar_{\incl}$ and $1/\wbar_{j}$.
}.
The two factors in Eq.~(\ref{eq:weight_incl}) may be interpreted in the following way:
The product of $w_{\incl}^{k}$ and the first factor, $w_{\incl}^{k} \, \frac{L \, \sigma_{\incl}}{\sum_{k=1}^{N_{\incl}} \, w_{\incl}^{k}}$,
corresponds to the weight that one would apply to an event in PS region $i$ 
in case no exclusive samples are available and the signal or background estimate in PS region $i$ is based solely on the inclusive sample.
The availability of the additional exclusive samples increases the number of simulated events in the PS region $i$, 
from $N_{\incl} \, P_{\incl}^{i}$ to $N_{\incl} \, P_{\incl}^{i} + \sum_{j} \, N_{j} \, P_{j}^{i}$,
and reduces the weights that are applied to simulated events falling into the region $i$.
The reduction in the event weight is given by the second factor in Eq.~(\ref{eq:weight_incl}).
It has the effect of reducing the statistical uncertainty on the signal or background estimate in PS region $i$
by the square-root of this factor,
\ie by $\sqrt{\frac{P_{\incl}^{i} \, \sum_{k=1}^{N_{\incl}} \, w_{\incl}^{k}}{P_{\incl}^{i} \, \sum_{k=1}^{N_{\incl}} \, w_{\incl}^{k} + \sum_{j} \, P_{j}^{i} \, \sum_{k=1}^{N_{j}} \, w_{j}^{k}}}$.

\section{Examples}
\label{sec:examples}

In this Section, we illustrate the formalism developed in Section~\ref{sec:stitching_weights} with concrete examples,
drawn from two different applications: the modelling of $\PW$+jets production in physics analyses at the LHC
and the estimation of trigger rates at the HL-LHC.

\subsection{Modelling of \texorpdfstring{$\PW$}{W}+jets production in physics analyses at the LHC}
\label{sec:examples_background_yield}

The production of $\PW$ bosons is interesting to study at the LHC for several reasons.
The measurement of the mass of the $\PW$ boson is an important input to global fits to SM parameters~\cite{Baak:2014ora}.
The fits allow to test the overall consistency of the SM and to set constraints on physics beyond the SM.
The sensitivity of these fits is currently limited by the precision of the $\PW$ boson mass measurement~\cite{Baak:2014ora}.
Differential measurements of the cross section for $\PW$+jets production 
are used to constrain parton distribution functions~\cite{CMS:2016qqr,ATLAS:2016nqi,ATLAS:2019fgb,CMS:2020cph}.
In particular, the measurement of the associated production of a $\PW$ boson with a charm quark
provides sensitivity to the strange quark content of the proton~\cite{CMS:2013wql,ATLAS:2014jkm,CMS:2018dxg} 
and allows to tune MC generators to improve the modelling of heavy flavour production at hadron colliders.
The production of $\PW$ bosons also constitutes a relevant background to measurements of other SM processes
and to searches for new physics, see for example Refs.~\cite{ATLAS:2014aga,Aad:2019yxi,CMS-HIG-13-027,CMS-HIG-17-006}.
In this section, we focus on $\PW$+jets production with subsequent leptonic decay of the $\PW$ boson.

Simulated samples of $\PW$+jets events have been produced for $\Pp\Pp$ collisions at $\sqrt{s}=13$~\TeV center-of-mass energy
using matrix elements computed at leading order (LO) accuracy in perturbative quantum chromodynamics (pQCD)
with the program \MGvATNLO $2.6.5$~\cite{MGvATNLO}.
The parton distribution functions of the proton are modeled using the NNPDF3.1 set~\cite{NNPDF:2017mvq}.
Parton showering, hadronization, and the underlying event are modeled using the program \PYTHIA $v8.240$~\cite{PYTHIA} with the tune \textrm{CP5}~\cite{Sirunyan:2019dfx}.
The matching of matrix elements to parton showers is done using the \textrm{MLM} scheme~\cite{Alwall:2007fs}.
We restrict the analysis of these samples to particles originating from the hard-scatter interaction and do not add any pileup to these samples.
Samples containing either $1$, $2$, $3$, or $4$ jets at matrix-element level are complemented by an ``inclusive'' sample 
and by samples binned in the scalar sum in $\pT$ of these jets.
We denote the multiplicity of jets at the matrix-element level by the symbol $N_{\jet}$ and the scalar sum in $\pT$ of these jets by the symbol $\HT$.
The inclusive and $\HT$-binned samples contain events with between $0$ and $4$ jets at the matrix-element level.

The weights $w_{j}^{k}$ and $w_{\incl}^{k}$ are equal to one for all events in these samples.
Thus, $\sum_{k=1}^{N_{j}} \, w_{j}^{k} = N_{j}$ and $\sum_{k=1}^{N_{\incl}} \, w_{\incl}^{k} = N_{\incl}$ for this example,
which allows us to simplify Eq.~(\ref{eq:weight_incl}) to:
\begin{equation}
s^{i} = \frac{L \, \sigma_{\incl}}{N_{\incl}} \, \frac{P_{\incl}^{i} \, N_{\incl}}{P_{\incl}^{i} \, N_{\incl} + P_{j}^{i} \, N_{j}} \, .
\label{eq:weight_incl_simplified}
\end{equation}

All samples are normalized using a $k$-factor of $1.14$, given by the ratio of the inclusive cross section computed at next-to-next-to leading order (NNLO) accuracy in pQCD,
with electroweak corrections taken into account up to NLO accuracy~\cite{Li:2012wna},
and the inclusive cross section computed at LO accuracy by the program \MGvATNLO.
The product of the inclusive $\PW$+jets production cross section times the branching fraction for the decay to a charged lepton and a neutrino amounts to $61.5$~nb.

We will demonstrate the stitching of these samples based on the two observables $N_{\jet}$ and $\HT$.
The PS region in which we perform the stitching will be either one- or two-dimensional.
We will show that for our formalism
it makes little difference whether the stitching is performed in one dimension or in two.
The stitching of $\PW$+jets samples based on the observable $N_{\jet}$ will be discussed first
and then we will discuss the stitching of $\PW$+jets samples based on the two observables $N_{\jet}$ and $\HT$.

\subsubsection{Stitching of \texorpdfstring{$\PW$}{W}+jets samples by \texorpdfstring{$N_{\jet}$}{Njet}}
\label{sec:WJets_vs_Njet}

In this example, an inclusive $\PW$+jets sample simulated at LO accuracy in pQCD 
is stitched with exclusive samples containing events with $N_{\jet}$ equal to either $1$, $2$, $3$, or $4$.
The inclusive sample contains events with $N_{\jet}$ between $0$ and $4$.
We partition the PS into slices based on the multiplicity of jets at the matrix-element level and set the index $i$ equal to $N_{\jet}$.
The number of events in each MC sample is chosen such that the stitching weights decrease by about a factor of two for each increase in jet multiplicity.
The decrease in the cross section as function of $N_{\jet}$
allows to reduce the statistical uncertainties in the tail of the $N_{\jet}$ distribution
without significantly increasing the expenditure of computing resources required to produce and store these samples.
The number of events contained in each sample and the values of the probabilities $P_{\incl}^{i}$ and $P_{j}^{i}$ are given in Table~\ref{tab:samples_and_probabilities_WJets_vs_Njet}.
The probabilities $P_{\incl}^{1}$, $P_{\incl}^{2}$, $P_{\incl}^{3}$, and $P_{\incl}^{4}$ are computed by taking the ratio of cross sections,
computed at LO accuracy by the program \MGvATNLO,
for the exclusive samples with respect to the cross section $\sigma_{\incl}$ of the inclusive sample.
The probability $P_{\incl}^{0}$ is obtained using the relation $P_{\incl}^{0} = 1 - \sum_{i=1}^{4} P_{\incl}^{i}$.
The probabilities $P_{j}^{i}$ for the exclusive samples are $1$ if $i=j$ and $0$ otherwise,
as each of the exclusive samples $j$ covers exactly one PS region $i$.
The corresponding stitching weights $s^{i}$, computed according to Eq.~(\ref{eq:weight_incl_simplified}), are given in Table~\ref{tab:weights_WJets_vs_Njet}.

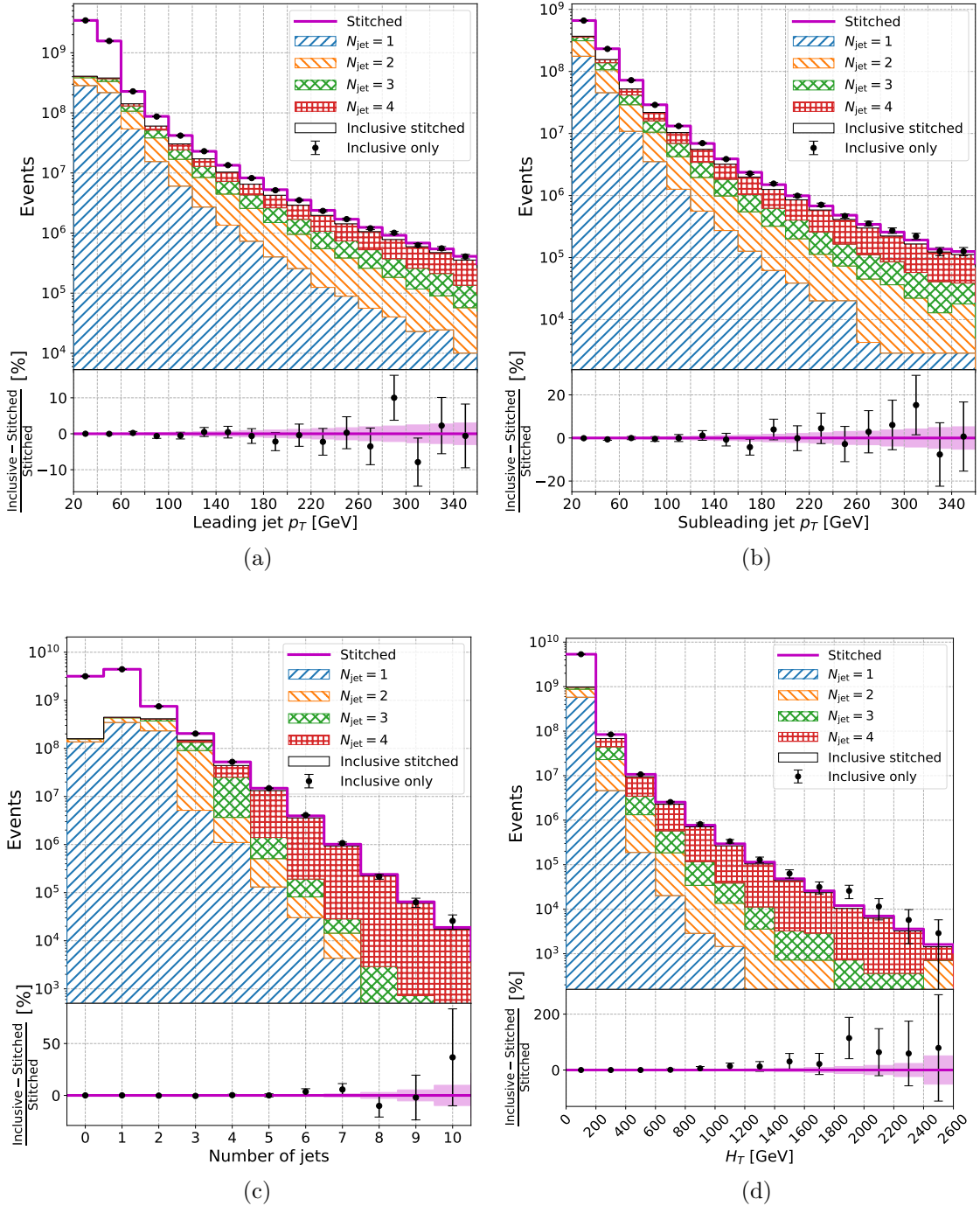
\begin{figure}
\setlength{\unitlength}{1mm}
\begin{center}
\begin{picture}(180,182)(0,0)
\put(6.5, 100.0){\mbox{\includegraphics*[height=82mm]{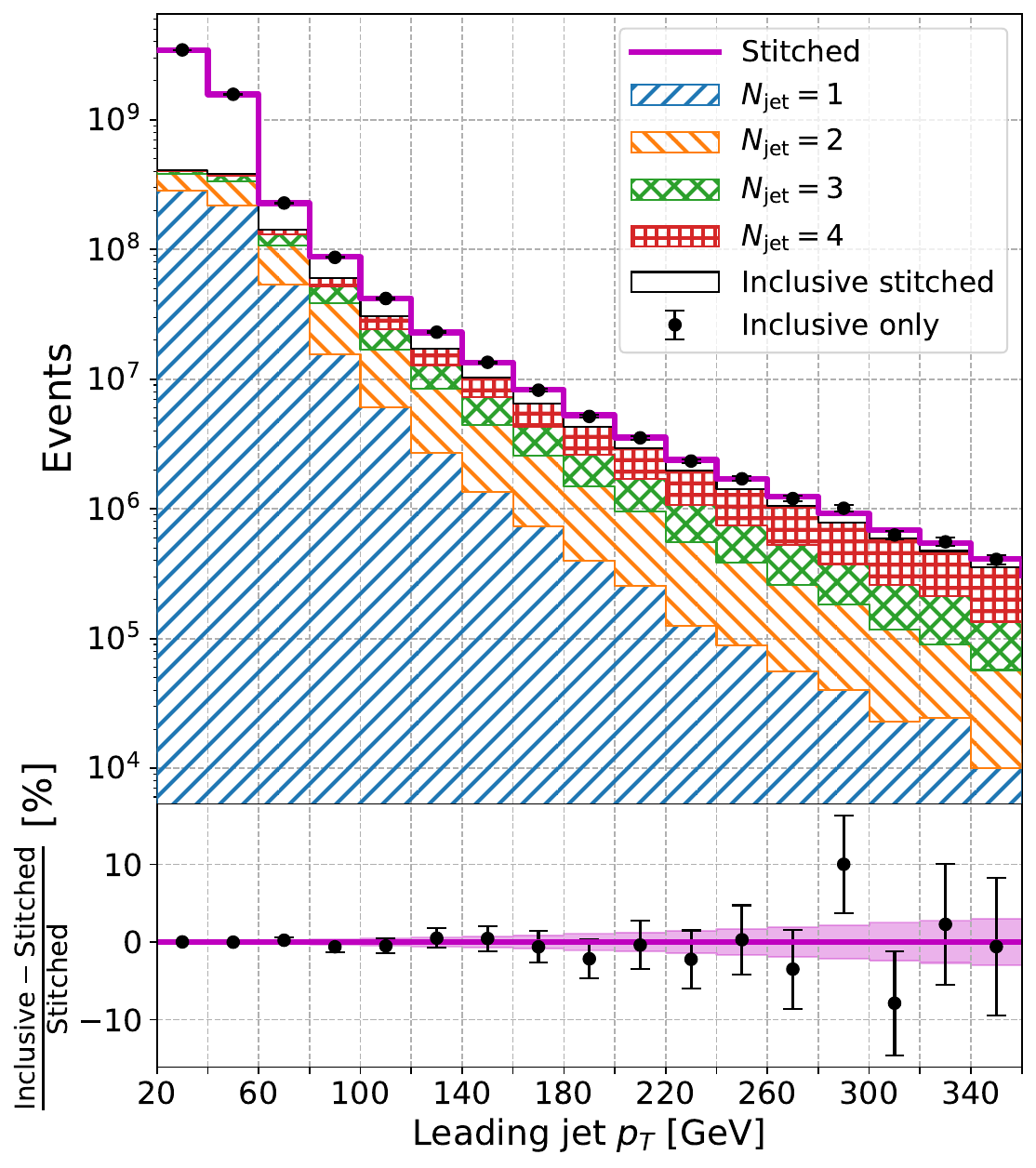}}}
\put(81.5, 100.0){\mbox{\includegraphics*[height=82mm]{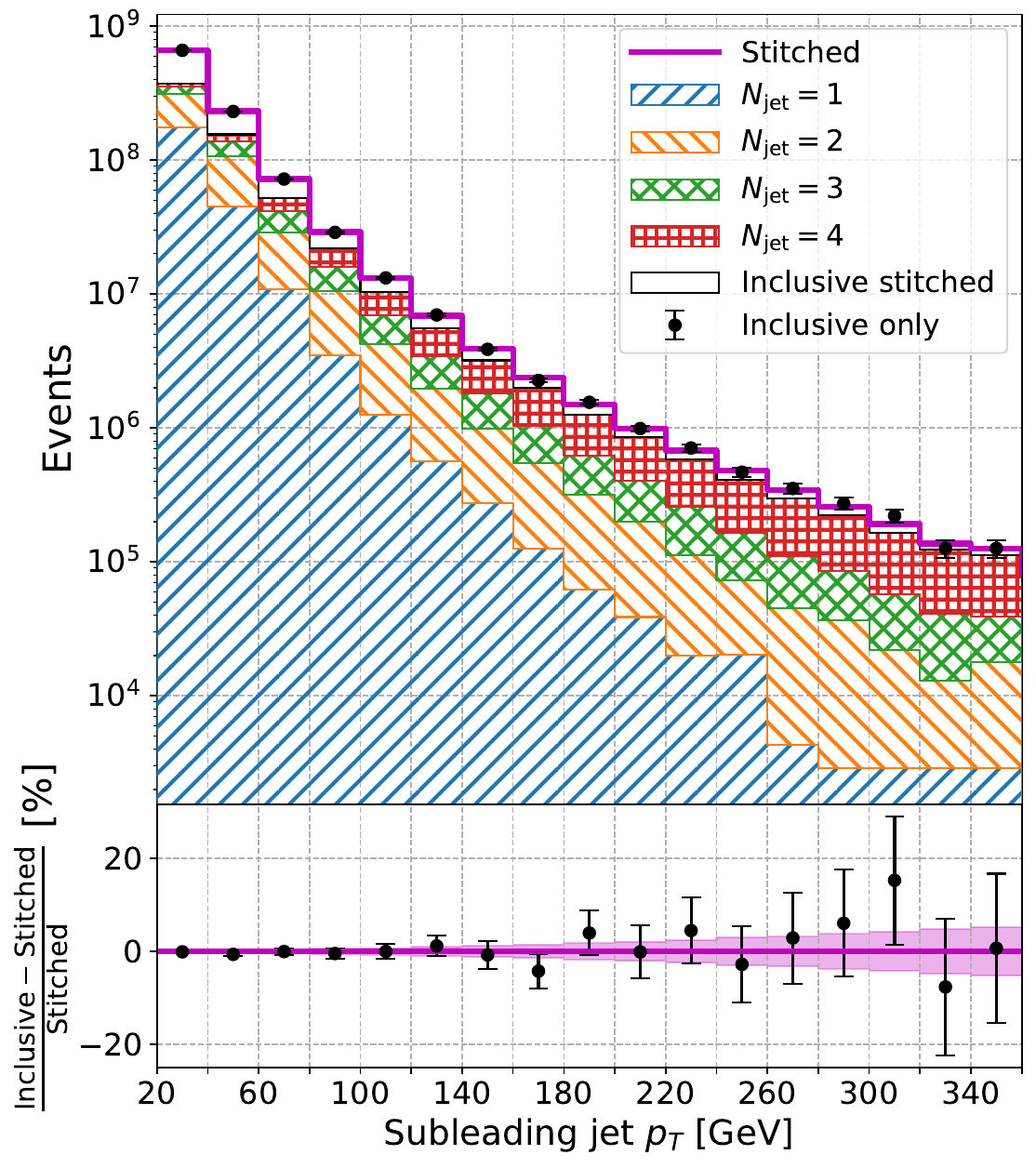}}}
\put(6.5, 4.0){\mbox{\includegraphics*[height=82mm]{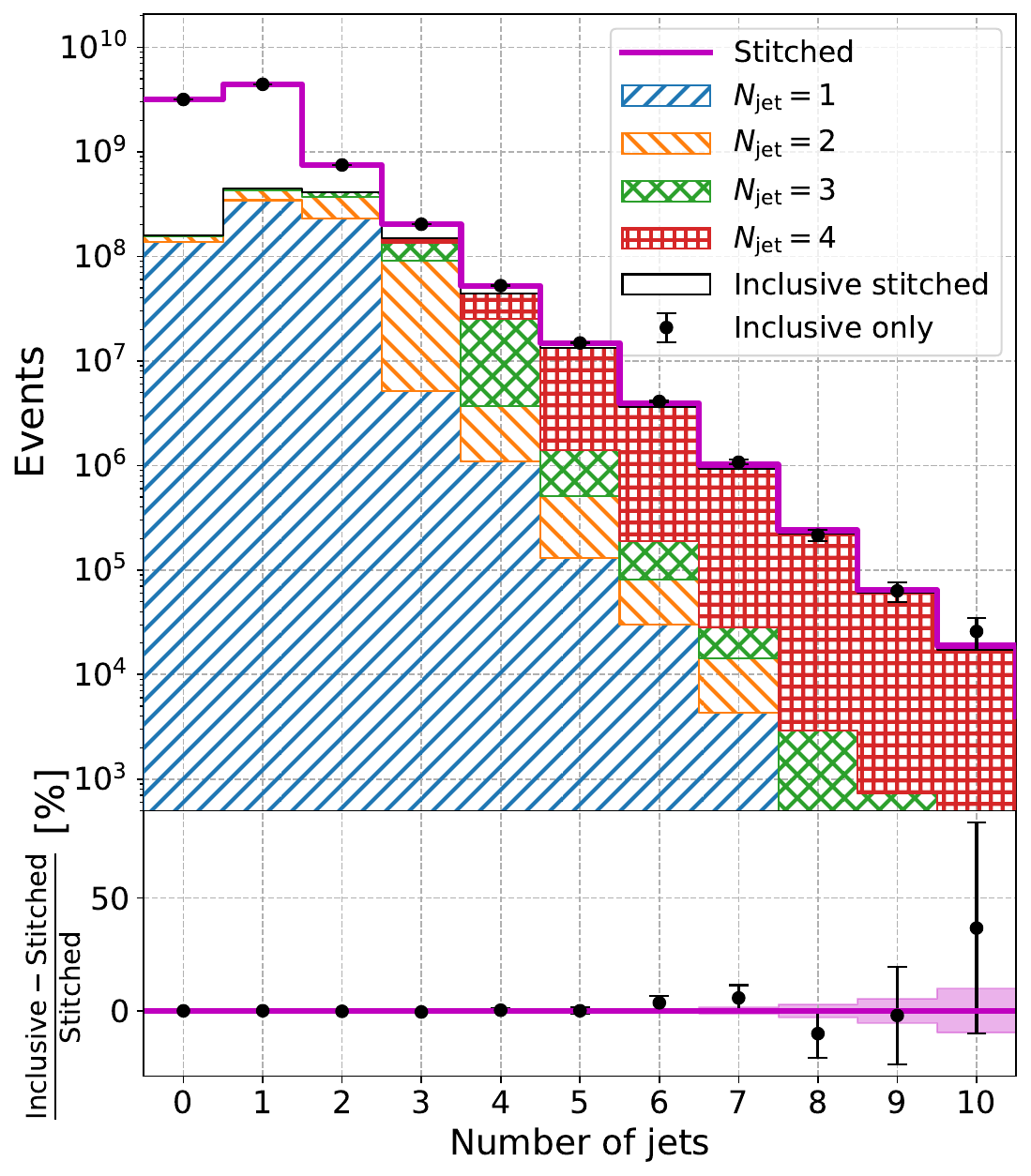}}}
\put(81.5, 4.0){\mbox{\includegraphics*[height=82mm]{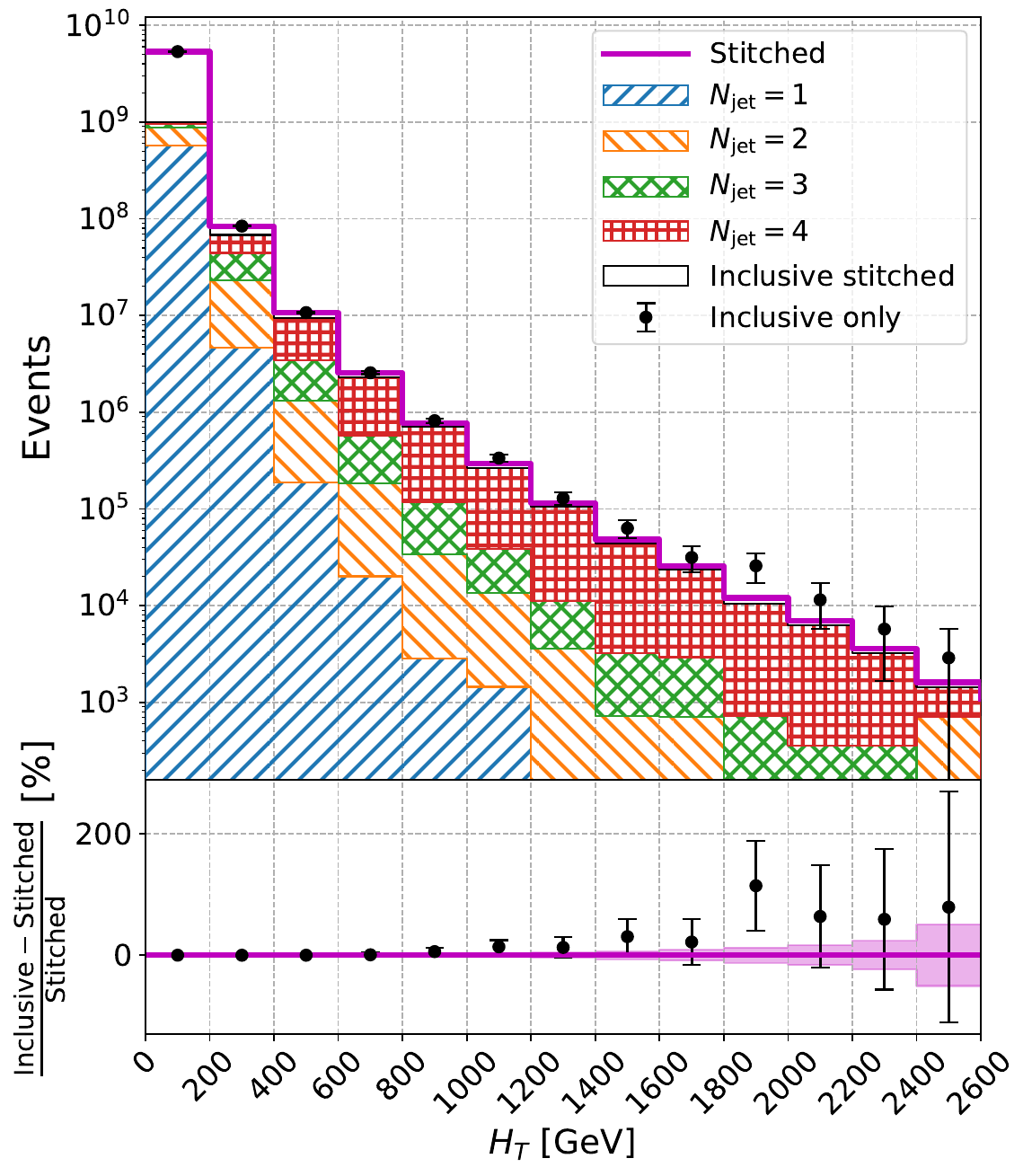}}}
\put(43.0, 96.0){\small (a)}
\put(118.0, 96.0){\small (b)}
\put(43.0, 0.0){\small (c)}
\put(118.0, 0.0){\small (d)}
\end{picture}
\end{center}
\caption{
  Distributions in $\pT$ of the (a) leading and (b) subleading jet,
  in (c) the multiplicity of generator-level jets and in (d) the observable $\HT$, the scalar sum in $\pT$ of these jets,
  for the case of $\PW$+jets samples that are stitched based on the observable $N_{\jet}$ at the matrix-element level.
  The $\PW$ bosons are required to decay leptonically.
  The event yields are computed for an integrated luminosity of $140\fbinv$.
}
\label{fig:controlPlots_WJets_vs_Njet}
\end{figure}

In order to demonstrate that the stitching procedure is unbiased,
we compare the normalization and shape of distributions obtained using the stitching procedure with the normalization and shape of distributions obtained from the inclusive sample.
Distributions in $\pT$ of the ``leading'' and ``subleading'' jet (the jets of, respectively, highest and second-highest $\pT$ in the event),
in the multiplicity of jets and in the observable $\HT$ are shown in Fig.~\ref{fig:controlPlots_WJets_vs_Njet}.
The distributions obtained from the inclusive sample are represented by black markers (``inclusive only''),
while those obtained by applying the stitching procedure to the combination of the inclusive sample and the samples binned in $N_{\jet}$ are represented by pink lines (``stitched'').
The contributions of individual exclusive samples $j$ to the stitched distribution are indicated by shaded areas of different color in the upper part of each figure.
The white area (``inclusive stitched'') represents the contribution of the inclusive sample to the stitched distribution.
The shaded areas of different color and the white area add up to the pink line.
We remark that the ``inclusive only'' and ``inclusive stitched'' distributions contain the exact same events. 
The sole difference between these two distributions 
is that the stitching weights, given in Table~\ref{tab:weights_WJets_vs_Njet},
are applied to the ``inclusive stitched'', but not to the ``inclusive only'' distribution.
In the lower part of each figure, we show the difference in normalization and shape between the distribution obtained using the stitching procedure
and the distribution obtained when using solely the inclusive sample.
The differences are given relative to the distribution obtained from our stitching procedure.
The size of statistical uncertainties on the ``inclusive only'' and ``stitched'' distributions
is visualized in the lower part of each figure and is represented by the length of the error bars and by the height of the dark shaded area, respectively.
The jets shown in the figure are reconstructed using the anti-\kt algorithm~\cite{Cacciari:2008gp,Cacciari:2011ma} with a distance parameter of $0.4$,
using all stable generator-level particles (after hadron shower and hadronization) except neutrinos as input, and are required to satisfy the selection criteria $\pT > 25$~\GeV and $\vert\eta\vert < 5.0$.
The observable $\HT$ is computed as the scalar sum in $\pT$ of these jets.
Note that the multiplicity of jets and the observable $\HT$ shown in Fig.~\ref{fig:controlPlots_WJets_vs_Njet} 
differ from the observables $N_{\jet}$ and $\HT$ that are used in the stitching procedure:
The former refer to jets at the generator (detector) level, while the latter refer to jets at the matrix-element level.
The distributions are normalized to an integrated luminosity of $140$~fb$^{-1}$.

The distributions for the inclusive sample and for the sum of inclusive plus exclusive samples, with the stitching weights applied, are in agreement within the statistical uncertainties.
The exclusive samples reduce the statistical uncertainties in particular in the tails of the distributions.

\begin{table}[h!]
\begin{center}
\def\arraystretch{1.3}
\begin{tabular}{l|c|c|c|ccccc}
\multirow{2}{20mm}{Sample} & Index & Number    & Cross                    & \multicolumn{5}{c}{Probabilities}               \\
                           & $j$   & of events & section [nb]$^{\dagger}$ & $P^{0}$ & $P^{1}$ & $P^{2}$ & $P^{3}$ & $P^{4}$ \\
\hline
Inclusive                  & $-$   & $3 \times 10^{6}$ & $61.5$ & $0.758$ & $0.167$ & $0.052$ & $0.015$ & $0.007$ \\
$N_{\jet} = 1$             & $1$   & $5 \times 10^{6}$   & $10.1$  & $0$     & $1$     & $0$     & $0$     & $0$  \\
$N_{\jet} = 2$             & $2$   & $4.7 \times 10^{6}$ & $3.21$  & $0$     & $0$     & $1$     & $0$     & $0$  \\
$N_{\jet} = 3$             & $3$   & $3.2 \times 10^{5}$ & $0.938$ & $0$     & $0$     & $0$     & $1$     & $0$  \\
$N_{\jet} = 4$             & $4$   & $3.3 \times 10^{5}$ & $0.443$ & $0$     & $0$     & $0$     & $0$     & $1$  \\
\end{tabular}
\end{center}
$^{\dagger}$ Computed at LO accuracy in pQCD, then scaled to NNLO
\caption{
  Number of events in the inclusive $\PW$+jets sample and in the $\PW$+jets samples produced in bins of $N_{\jet}$,
  corresponding cross sections,
  and probabilities $P^{i}$ for the events in the inclusive and exclusive samples to populate the different PS regions $i$.
}
\label{tab:samples_and_probabilities_WJets_vs_Njet}
\end{table}

\begin{table}[h!]
\begin{center}
\begin{tabular}{l|ccccc}
 & \multicolumn{5}{c}{Multiplicity of jets} \\
 & $0$ & $1$ & $2$ & $3$ & $4$ \\
\hline
Stitching weight & $2870$ & $1440$ & $714$ & $362$ & $180$ \\
\end{tabular}
\end{center}
\caption{
  Stitching weights $s^{i}$ for the case that the inclusive and exclusive $\PW$+jets samples 
  given in Table~\ref{tab:samples_and_probabilities_WJets_vs_Njet}
  are stitched based on $N_{\jet}$.
  The weights are computed for an integrated luminosity of $140\fbinv$.
}
\label{tab:weights_WJets_vs_Njet}
\end{table}

\subsubsection{Stitching of \texorpdfstring{$\PW$}{W}+jets samples by \texorpdfstring{$N_{\jet}$}{Njet} and \texorpdfstring{$\HT$}{HT}}
\label{sec:WJets_vs_Njet_and_HT}

This example extends the previous example.
It demonstrates the stitching procedure based on two observables, $N_{\jet}$ and $\HT$.
The exclusive samples are simulated for jet multiplicities of $N_{\jet} = 1$, $2$, $3$, and $4$ 
and for $\HT$ in the ranges $70$-$100$, $100$-$200$, $200$-$400$, $400$-$600$, $600$-$800$, $800$-$1200$, $1200$-$2500$, and $> 2500$~\GeV (up to the kinematic limit).
We refer to the exclusive samples produced in slices of $N_{\jet}$ as the ``$N_{\jet}$-samples''
and to the samples simulated in ranges in $\HT$ as the ``$\HT$-samples''.
The inclusive sample contains events with jet multiplicities between $0$ and $4$ and covers the full range in $\HT$.
The number of events in the $\HT$-samples are given in Table~\ref{tab:samples_WJets_vs_Njet_and_HT}.
The information for the inclusive sample and for the $N_{\jet}$-samples is the same as for the previous example
and is given in Table~\ref{tab:samples_and_probabilities_WJets_vs_Njet}.

\begin{figure}
\setlength{\unitlength}{1mm}
\begin{center}
\begin{picture}(180,82)(0,0)
\includegraphics*[height=82mm]{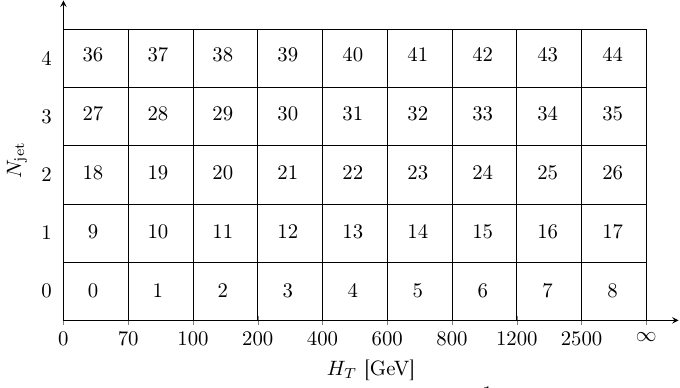}
\end{picture}
\end{center}
\caption{
  Definition of the PS regions $i$ in the plane of $N_{\jet}$ versus $\HT$,
  for the case of $\PW$+jets samples that are stitched based on the observables $N_{\jet}$ and $\HT$.
}
\label{fig:regions_WJets_vs_Njet_and_HT}
\end{figure}

The corresponding PS regions $i$, defined in the plane of $N_{\jet}$ versus $\HT$, are shown in Fig.~\ref{fig:regions_WJets_vs_Njet_and_HT}.
In total, the probabilities $P_{\incl}^{i}$ and $P_{j}^{i}$ and the corresponding stitching weights $s^{i}$ are computed for $45$ separate PS regions.

In some of the $45$ PS regions, the probabilities $P_{\incl}^{i}$ are rather low, on the level of $10^{-7}$.
In order to reduce the statistical uncertainties on the probabilities $P_{\incl}^{i}$ and $P_{j}^{i}$,
we compute these probabilities using the following procedure:
For all regions $i$ of PS that are covered by the inclusive sample and by one or more $N_{\jet}$- or $\HT$-samples,
we determine the probabilities $P_{\incl}^{i}$ and $P_{j}^{i}$ by the method of least squares~\cite{Cowan:1998ji}.
Details of the computation are given in the appendix.
The probability $P_{\incl}^{0}$ for the PS region $N_{\jet} = 0$ and $\HT < 70$~\GeV,
which is solely covered by the inclusive sample,
is computed according to the relation $P_{\incl}^{0} = 1 - \sum_{i=1}^{44} \, P_{\incl}^{i}$.
The numerical values of the probabilities $P_{\incl}^{i}$ and $P_{j}^{i}$ obtained by the least-square method
and of the stitching weights $s^{i}$, computed according to Eq.~(\ref{eq:weight_incl_simplified}), are given in Tables~\ref{tab:probabilities_WJets_vs_Njet_and_HT}
and~\ref{tab:weights_WJets_vs_Njet_and_HT}.

\begin{figure}
\setlength{\unitlength}{1mm}
\begin{center}
\begin{picture}(180,182)(0,0)
\put(6.5, 100.0){\mbox{\includegraphics*[height=82mm]{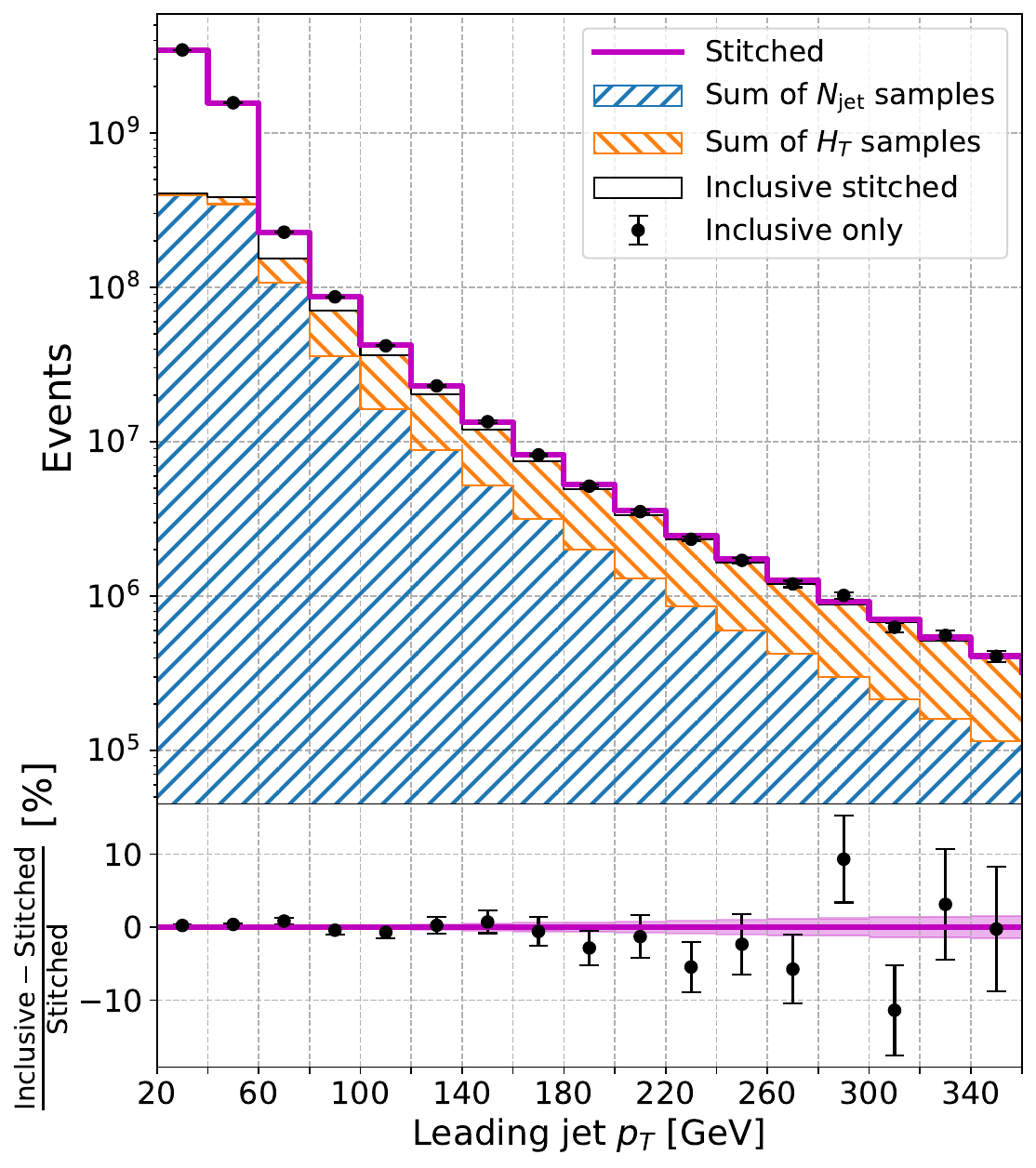}}}
\put(81.5, 100.0){\mbox{\includegraphics*[height=82mm]{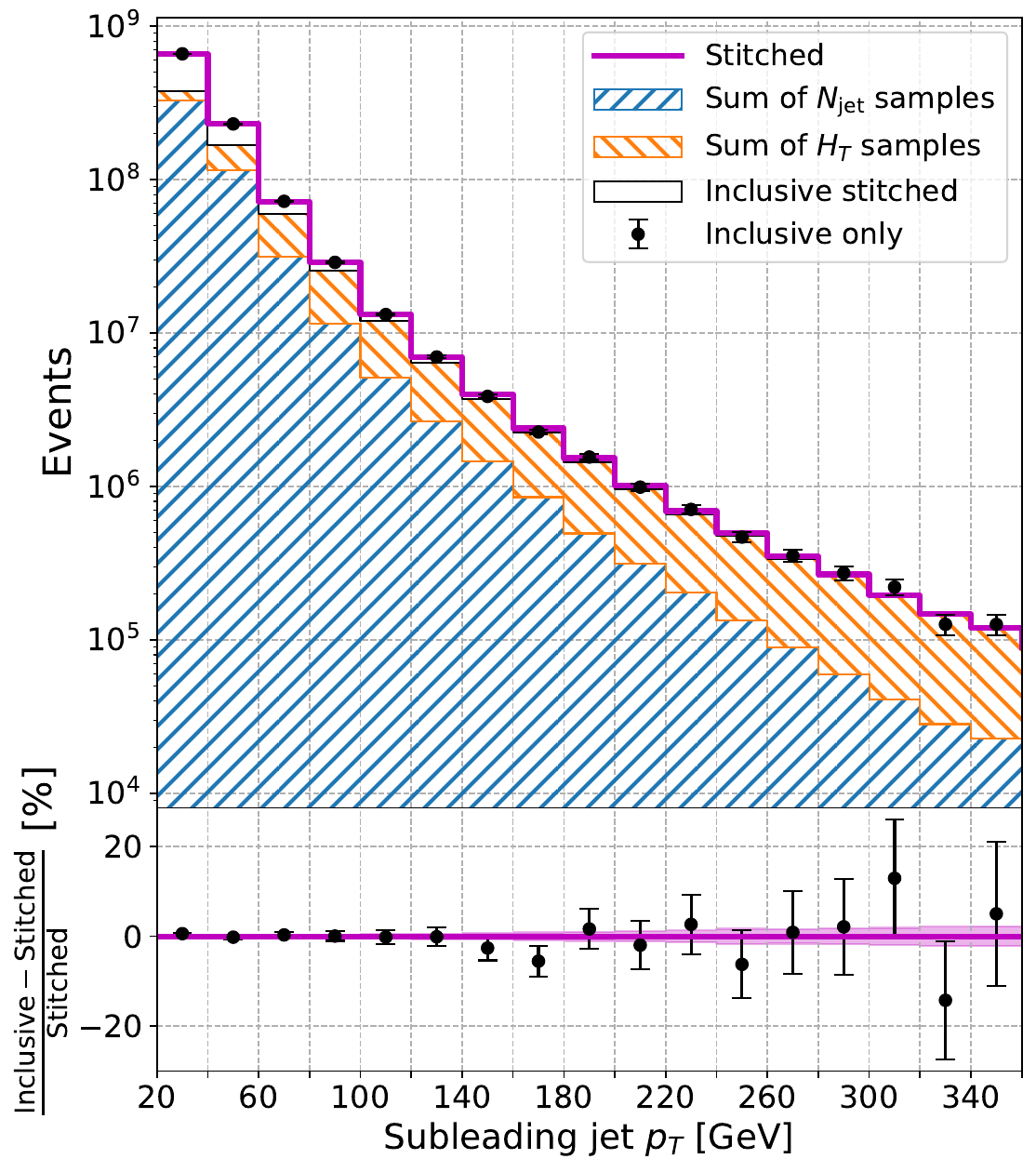}}}
\put(6.5, 4.0){\mbox{\includegraphics*[height=82mm]{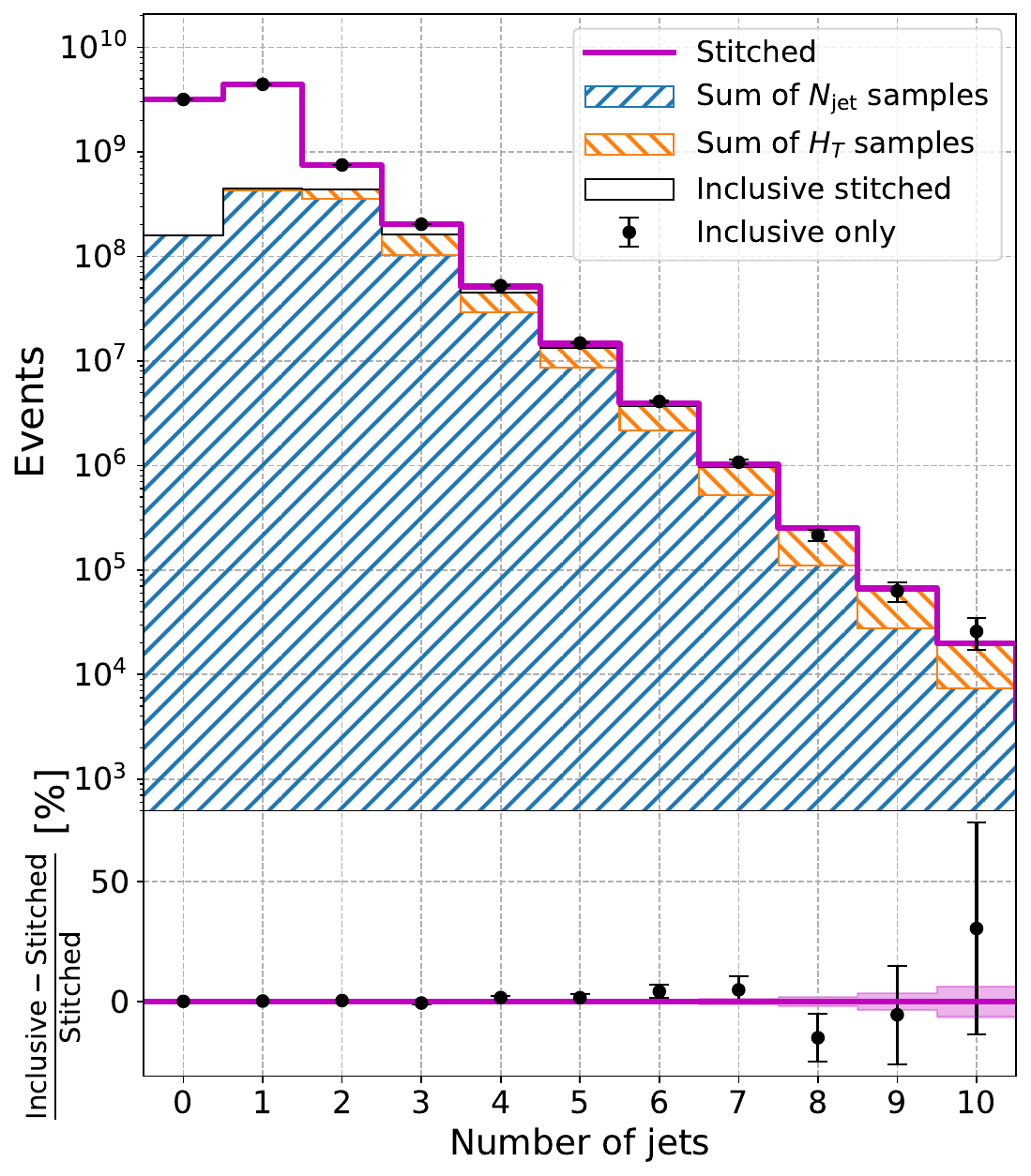}}}
\put(81.5, 4.0){\mbox{\includegraphics*[height=82mm]{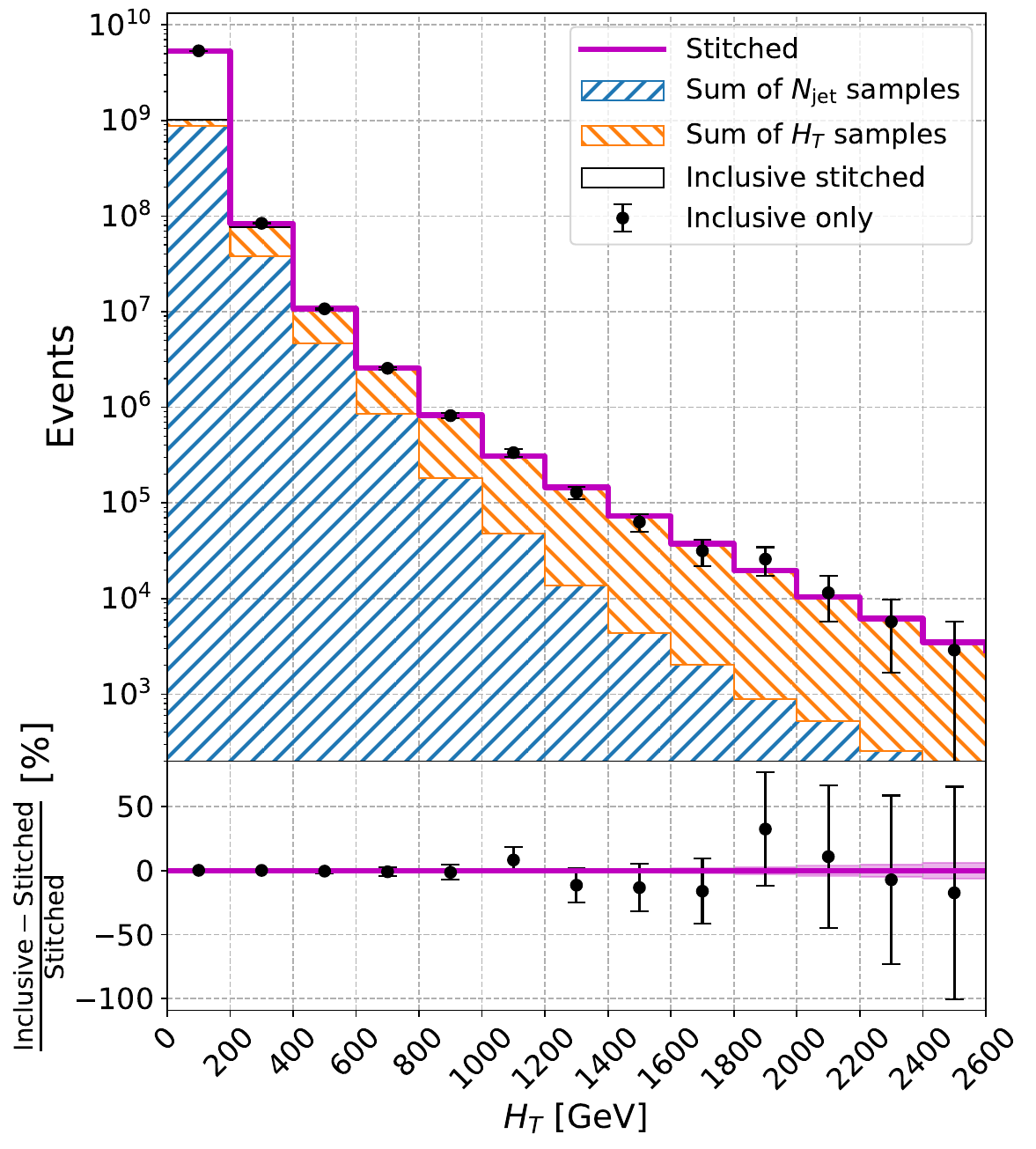}}}
\put(43.0, 96.0){\small (a)}
\put(118.0, 96.0){\small (b)}
\put(43.0, 0.0){\small (c)}
\put(118.0, 0.0){\small (d)}
\end{picture}
\end{center}
\caption{
  Distributions in $\pT$ of the (a) leading and (b) subleading jet,
  in (c) the multiplicity of generator-level jets and in (d) the observable $\HT$, the scalar sum in $\pT$ of these jets,
  for the case of $\PW$+jets samples that are stitched based on the observables $N_{\jet}$ and $\HT$ at the matrix-element level.
  The $\PW$ bosons are required to decay leptonically.
  The event yields are computed for an integrated luminosity of $140\fbinv$.
}
\label{fig:controlPlots_WJets_vs_Njet_and_HT}
\end{figure}

Distributions in $\pT$ of the leading and subleading jet,
in the multiplicity of jets, and in the observable $\HT$ 
obtained from our stitching procedure are compared to the distributions obtained from the inclusive sample in Fig.~\ref{fig:controlPlots_WJets_vs_Njet_and_HT}.
The distributions obtained by stitching the inclusive sample with the samples binned in $N_{\jet}$ and in $\HT$ are represented by pink lines,
while those obtained when using solely the inclusive sample are represented by black markers.
The weights given in Table~\ref{tab:weights_WJets_vs_Njet_and_HT} are applied to the stitched distributions.
The contribution of all $N_{\jet}$-binned samples to the stitched distribution is represented by the blue shaded area (``sum of $N_{\jet}$ samples''),
while the contribution of all samples produced in ranges in $\HT$ is represented by the yellow shaded area (``sum of $\HT$ samples'') in the upper part of the figures.
Following Fig.~\ref{fig:controlPlots_WJets_vs_Njet},
the white area represents the contribution of the inclusive sample to the stitched distribution.
The jets are reconstructed as described in Section~\ref{sec:WJets_vs_Njet} and are required to pass the selection criteria $\pT > 25$~\GeV and $\vert\eta\vert < 5.0$.
In the lower part of each figure, we again show the difference between the distributions obtained from the inclusive sample and obtained by using our stitching procedure,
and also the respective statistical uncertainties.
As one would expect, the addition of samples simulated in ranges in $\HT$ to the example given in Section~\ref{sec:WJets_vs_Njet}
reduces the statistical uncertainties in the tails of the distributions.
The reduction is most pronounced in the tails of the distributions in leading and subleading jet $\pT$ and in the observable $\HT$.

\begin{table}[h!]
\begin{center}
\def\arraystretch{1.3}
\begin{tabular}{l|c|c|c}
\multirow{2}{20mm}{Sample}       & Index & Number    & Cross                    \\
                                 & $j$   & of events & section [pb]$^{\dagger}$ \\
\hline
$  70 \leqslant \HT <  100$~\GeV &  $5$  & $1.4 \times 10^{5}$ & $1430$  \\
$ 100 \leqslant \HT <  200$~\GeV &  $6$  & $2.8 \times 10^{5}$ & $1410$  \\
$ 200 \leqslant \HT <  400$~\GeV &  $7$  & $1.5 \times 10^{5}$ & $379$   \\
$ 400 \leqslant \HT <  600$~\GeV &  $8$  & $  4 \times 10^{4}$ & $51.3$  \\
$ 600 \leqslant \HT <  800$~\GeV &  $9$  & $  2 \times 10^{4}$ & $12.6$  \\
$ 800 \leqslant \HT < 1200$~\GeV & $10$  & $  2 \times 10^{4}$ & $5.34$  \\
$1200 \leqslant \HT < 2500$~\GeV & $11$  & $1.3 \times 10^{4}$ & $1.51$  \\
$       \HT \geqslant 2500$~\GeV & $12$  & $6.5 \times 10^{2}$ & $0.052$ \\
\end{tabular}
\end{center}
$^{\dagger}$ Computed at LO accuracy in pQCD, then scaled to NNLO
\caption{
  Number of events in the $\PW$+jets samples produced in ranges in $\HT$ and corresponding cross sections.
}
\label{tab:samples_WJets_vs_Njet_and_HT}
\end{table}

\subsection{Estimation of trigger rates at the HL-LHC}
\label{sec:examples_trigger_rate}

We choose the task of estimating trigger rates for the upcoming high-luminosity data-taking period of the LHC as second example to illustrate the stitching procedure.
The ``rate'' of a trigger corresponds to the number of $\Pp\Pp$ collision events that satisfy the trigger condition per unit of time.
The estimation of trigger rates constitutes an important task for demonstrating the physics potential of the HL-LHC.
The HL-LHC physics program demands a large amount of integrated luminosity to be delivered by the LHC, 
in order to facilitate measurements of rare signal processes
(such as the precise measurement of $\PHiggs$ boson couplings and the study of $\PHiggs$ boson pair production),
as well as to enhance the sensitivity of searches for new physics, by the ATLAS and CMS experiments.
In order to satisfy this demand, the HL-LHC is expected to operate at an instantaneous luminosity of $5$-$7.5 \times 10^{34}$~cm$^{-2}$~s$^{-1}$
at a center-of-mass energy of $\sqrt{s} = 14$~\TeV~\cite{TDR_Phase2_LHC}.
The challenge of developing triggers for the HL-LHC is to design the triggers such that rare signal processes pass the triggers with a high efficiency,
while the rate of background processes gets reduced by many orders of magnitude, in order not to exceed bandwidth limitations on the detector read-out 
and on the rate with which events can be written to permanent storage.

The inelastic $\Pp\Pp$ scattering cross section at $\sqrt{s} = 14$~\TeV amounts to $\approx 80$~mb,
resulting in up to $200$ simultaneous $\Pp\Pp$ interactions per crossing of the proton beams at the nominal HL-LHC instantaneous luminosity~\cite{TDR_Phase2_LHC}.
The vast majority of these interactions are inelastic $\Pp\Pp$ scatterings with low momentum exchange,
which predominantly arise from the exchange of gluons between the colliding protons.
We refer to inelastic $\Pp\Pp$ scattering interactions with no further selection applied as ``minimum bias'' events.
In order to estimate the rates of triggers at the HL-LHC,
MC samples of minimum bias events are produced at LO in pQCD using the program \PYTHIA.
The minimum bias samples are complemented by samples of inelastic $\Pp\Pp$ scattering interactions
in which a significant amount of transverse momentum, denoted by the symbol $\pThat$, is exchanged between the scattered protons.
The stitching of the minimum bias samples with samples generated for different ranges in $\pThat$ allows to estimate the trigger rates with lower statistical uncertainties.

The production of MC samples used for estimating trigger rates at the HL-LHC
proceeds by first simulating one ``hard-scatter'' (HS) interaction within a given range in $\pThat$
and then adding a number of additional inelastic $\Pp\Pp$ scattering interactions of the minimum bias kind to the same event,
in order to simulate the pileup (PU).
We use the symbol $N_{\pileup}$ to denote the total number of PU interactions 
that occur in the same crossing of the proton beams as the HS interaction.
No selection on $\pThat$ is applied when simulating the PU interactions.
We remark that the distinction between the HS interaction and the PU interactions is artificial and solely made for the purpose of MC production.
The HS interaction and the PU interactions will be indistinguishable in the data that will be recorded at the HL-LHC:
The scattering in which the transverse momentum exchange between the protons amounts to $\pThat$ may occur in any of the $N_{\pileup} + 1$ simultaneous $\Pp\Pp$ interactions.
Our formalism treats the HS interaction and the PU interactions on an equal footing.

The ``inclusive'' sample in this example are events containing $N_{\pileup} + 1$ minimum bias interactions,
where for each event the number of PU interactions, $N_{\pileup}$, is sampled at random from the Poisson probability distribution:
\begin{equation}
\Poisson(N_{\pileup} \vert \Nbar) = \frac{\Nbar^{N_{\pileup}} \, e^{-\Nbar}}{N_{\pileup}!}
\label{eq:Poisson}
\end{equation}
with a mean $\Nbar = 200$.
The exclusive samples contain one HS interaction of transverse momentum within a specified range in $\pThat$ in addition to $N_{\pileup}$ minimum bias interactions.
The latter represent the PU.
The number $N_{\pileup}$ of PU interactions is again sampled at random from a Poisson distribution with a mean of $\Nbar = 200$.

We enumerate the ranges in $\pThat$ by the index $i$ and denote the number of $\pThat$ ranges used to produce the exclusive samples by the symbol $m$.
We further introduce the symbol $n_{i}$ to refer to the number of inelastic $\Pp\Pp$ scattering interactions that fall into the $i$-th interval in $\pThat$.
The inelastic $\Pp\Pp$ scatterings may occur either in the HS interaction or in any of the $N_{\pileup}$ PU interactions.
The ``phase space'' corresponding to a given event is represented by a vector $I=n_{1},\dots,n_{m}$ of dimension $m$.
The $i$-th component of this vector indicates the number of inelastic $\Pp\Pp$ scattering interactions that fall into the $i$-th interval in $\pThat$.

The probability $P^{I}$ for an event in the inclusive sample that contains $N_{\pileup}$ pileup interactions
to feature $n_{1}$ inelastic $\Pp\Pp$ scatterings that fall into the first interval in $\pThat$, $n_{2}$ that fall into the second,$\dots$, and $n_{m}$ that fall into the $m$-th 
follows a multinomial distribution~\cite{evans2011statistical} and is given by:
\begin{equation}
P_{\incl}^{I} = \frac{(N_{\pileup} + 1)!}{n_{1}! \, \dots \, n_{m}!} \, p_{1}^{n_{1}} \, \dots \, p_{m}^{n_{m}} \, ,
\label{eq:P_inclusive}
\end{equation}
where the symbols $p_{i}$ correspond to the probability for a single inelastic $\Pp\Pp$ scattering interaction to feature a transverse momentum exchange that falls into the $i$-th interval in $\pThat$.
The $n_{i}$ satisfy the condition $\sum_{i=1}^{m} \, n_{i} = N_{\pileup} + 1$.

The corresponding probability $P_{j}^{I}$ for an event in the $j$-th exclusive sample that contains $N_{\pileup}$ pileup interactions is given by:
\begin{equation}
P_{j}^{I} = \begin{cases}
\frac{N_{\pileup}!}{n_{1}! \, \dots \, (n_{j} - 1)! \, \dots \, n_{m}!} \, p_{1}^{n_{1}} \dots \, p_{j}^{(n_{j} - 1)} \dots \, p_{m}^{n_{m}} \, ,
  & \text{if $n_{j} \geq 1$} \\
0 \, , & \text{otherwise} \, .
\end{cases}
\label{eq:P_exclusive}
\end{equation}
The $n_{i}$ again satisfy the condition $N_{\pileup} + 1 = \sum_{i=1}^{m} \, n_{i}$.
The fact that for all events in the $j$-th exclusive sample the transverse momentum $\pThat$ that is exchanged in the HS interaction falls into the $j$-th interval in $\pThat$
implies that $N_{\pileup} + 1$ needs to be replaced by $N_{\pileup}$ and $n_{j}$ by $n_{j} - 1$ in Eq.~(\ref{eq:P_exclusive}) compared to Eq.~(\ref{eq:P_inclusive}),
as one of the inelastic $\Pp\Pp$ scatterings that fall into the $j$-th interval in $\pThat$ is ``fixed'' and thus not subject to the random fluctuations, which are modeled by the multinomial distribution.
The ratio of Eq.~(\ref{eq:P_exclusive}) to Eq.~(\ref{eq:P_inclusive}) is given by the expression:
\begin{equation}
\frac{P_{j}^{I}}{P_{\incl}^{I}} = \frac{n_{j}}{(N_{\pileup} + 1) \, p_{j}} \, .
\label{eq:P_ratio}
\end{equation}
The validity of Eq.~(\ref{eq:P_ratio}) includes the case $n_{j} = 0$.

The expression for the stitching weight $s^{I}$ is given by an expression similar to Eq.~(\ref{eq:weight_incl}),
the main difference being that the index $i$ is replaced by the vector $I$,
the probabilities $P_{\incl}^{i}$ and $P_{j}^{i}$ are replaced by the probabilities $P_{\incl}^{I}$ and $P_{j}^{I}$
and the product of luminosity times cross section, $L \, \sigma_{\incl}$, is replaced by the frequency $F$ of $\Pp\Pp$ collisions:
\begin{equation}
s^{I} = \frac{F}{\sum_{k=1}^{N_{\incl}} \, w_{\incl}^{k}} \, \frac{P_{\incl}^{I} \, \sum_{k=1}^{N_{\incl}} \, w_{\incl}^{k}}{P_{\incl}^{I} \, \sum_{k=1}^{N_{\incl}} \, w_{\incl}^{k} + \sum_{j} \, P_{j}^{I} \, \sum_{k=1}^{N_{j}} \, w_{j}^{k}} \, .
\label{eq:weight_tmp}
\end{equation}
The probabilities $P_{\incl}^{I}$ and $P_{j}^{I}$ are given by Eqs.(~\ref{eq:P_inclusive}) and~(\ref{eq:P_exclusive}).
Dividing both numerator and denominator on the right-hand side of Eq.~(\ref{eq:weight_tmp}) by $P_{\incl}^{I}$ and replacing the ratio $P_{j}^{I}/P_{\incl}^{I}$ by Eq.~(\ref{eq:P_ratio}) yields:
\begin{equation}
s^{I} = \frac{F}{\sum_{k=1}^{N_{\incl}} \, w_{\incl}^{k} + \sum_{j} \, \frac{n_{j}}{(N_{\pileup} + 1) \, p_{j}} \, \sum_{k=1}^{N_{j}} \, w_{j}^{k}} \, .
\label{eq:weight_trigger_rate}
\end{equation}
The sum over $j$ refers to the exclusive samples.
At the HL-LHC, the $\Pp\Pp$ collision frequency $F$ amounts to $28$~MHz~\footnote{
  The beams cross every $25$~ns, but $\Pp\Pp$ collisions occur only in $\approx 70\%$ of those beam crossings~\cite{TDR_Phase2_LHC}.}.
Eq.~(\ref{eq:weight_trigger_rate}) represents the equivalent of Eq.~(\ref{eq:weight_incl}),
tailored to the case of estimating trigger rates instead of estimating event yields.
The weights $w_{\incl}^{k}$ and $w_{j}^{k}$ are equal to one for all events in this example,
which allows to simplify Eq.~(\ref{eq:weight_trigger_rate}).
Using the relations $\sum_{k=1}^{N_{\incl}} \, w_{\incl}^{k} = N_{\incl}$ and $\sum_{k=1}^{N_{j}} \, w_{j}^{k} = N_{j}$,
we obtain the expression:
\begin{equation}
s^{I} = \frac{F}{N_{\incl} + \sum_{j} \, \frac{n_{j}}{(N_{\pileup} + 1) \, p_{j}} \, N_{j}} \, .
\label{eq:weight_trigger_rate_simplified}
\end{equation}

The ranges in $\pThat$ used to produce the exclusive samples and the number of events contained in each sample
are given in Table~\ref{tab:samples_trigger_rate}.
The association of the index $i$ to the different ranges in $\pThat$ and the 
corresponding values of the probabilities $p_{i}$ are given in Table~\ref{tab:p_trigger_rate}.
The probabilities $p_{i}$ are computed by taking the ratio of cross sections computed by the program \PYTHIA
for the case of single inelastic $\Pp\Pp$ scattering interactions with a transverse momentum exchange that is within the $i$-th interval in $\pThat$
and for the case that no condition is imposed on $\pThat$.

\begin{table}[h!]
\begin{center}
\begin{tabular}{l|c}
Sample                            & Number of events \\
\hline
Inclusive                         & $8 \times 10^{5}$ \\
$ 30 \leqslant \pThat <  50$~\GeV & $4 \times 10^{5}$ \\
$ 50 \leqslant \pThat <  80$~\GeV & $2 \times 10^{5}$ \\
$ 80 \leqslant \pThat < 120$~\GeV & $1 \times 10^{5}$ \\
$120 \leqslant \pThat < 170$~\GeV & $5 \times 10^{4}$ \\
$170 \leqslant \pThat < 300$~\GeV & $5 \times 10^{4}$ \\
$300 \leqslant \pThat < 470$~\GeV & $5 \times 10^{4}$ \\
$470 \leqslant \pThat < 600$~\GeV & $5 \times 10^{4}$ \\
$\pThat \geqslant 600$~\GeV       & $5 \times 10^{4}$ \\
\end{tabular}
\end{center}
\caption{
  Number of events in the inclusive and exclusive samples used to estimate trigger rates at the HL-LHC.
}
\label{tab:samples_trigger_rate}
\end{table}

\begin{table}[h!]
\begin{center}
\small
\begin{minipage}{16cm}
\begin{tabular}{l|cccccc}
Range in $\pThat$ [\GeV] & $< 30$ & $30$-$50$ & $50$-$80$ & $80$-$120$ & $120$-$170$ & $170$-$300$ \\
Index $i$           & $0$ & $1$ & $2$ & $3$ & $4$ & $5$ \\
\hline
Probability $p_{i}$ & $0.998$ & $1.51 \times 10^{-3}$ & $2.25 \times 10^{-4}$ & $3.38 \times 10^{-5}$ & $6.00 \times 10^{-6}$ & $1.55 \times 10^{-6}$ \\
\end{tabular}

\vspace{2mm}

\begin{tabular}{l|ccc}
Range in $\pThat$ [\GeV] & $300$-$470$ & $470$-$600$ & $> 600$ \\
Index $i$           & $6$ & $7$ & $8$ \\
\hline
Probability $p_{i}$ & $1.05 \times 10^{-7}$ & $8.73 \times 10^{-9}$ & $3.12 \times 10^{-9}$ \\
\end{tabular}
\end{minipage}
\end{center}
\caption{
  Probabilities $p_{i}$ for a single inelastic $\Pp\Pp$ scattering interaction to feature a transverse momentum exchange 
  between the protons that is within the $i$-th interval in $\pThat$.
}
\label{tab:p_trigger_rate}
\end{table}

\begin{figure}
\setlength{\unitlength}{1mm}
\begin{center}
\includegraphics*[height=76mm]{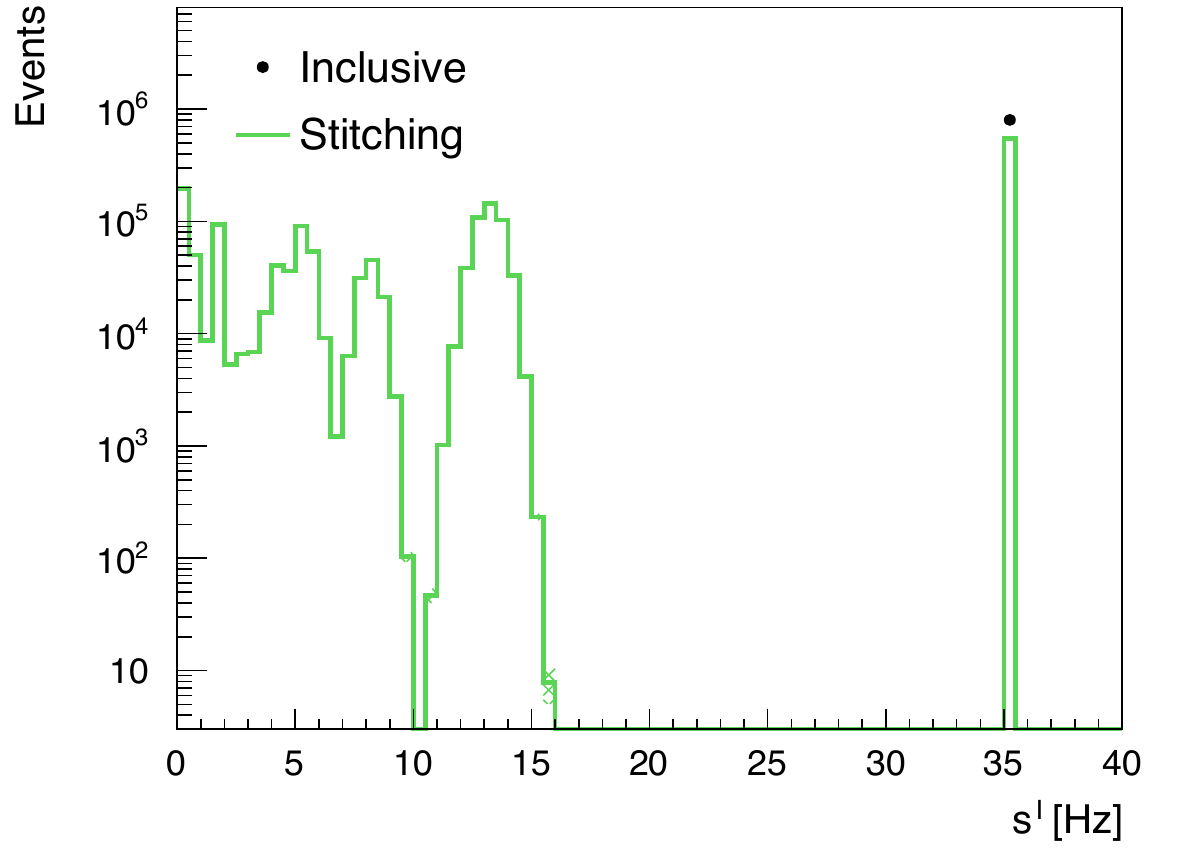}
\end{center}
\caption{
  Stitching weights $s^{I}$, computed according to Eq.~(\ref{eq:weight_trigger_rate_simplified}), 
  for the inclusive sample and for the samples produced in ranges of $\pThat$.
}
\label{fig:weight_trigger_rate}
\end{figure}

We cannot give numerical values of the stitching weights $s^{I}$ for this example,
as $I$ is a high-dimensional vector, and also because the stitching weights vary depending on $N_{\pileup}$.
Instead, we show in Fig.~\ref{fig:weight_trigger_rate} the spectrum of the stitching weights
that we obtain when inserting the numbers given in Tables~\ref{tab:samples_trigger_rate} and~\ref{tab:p_trigger_rate} into Eq.~(\ref{eq:weight_trigger_rate_simplified}).
For comparison, we also show the corresponding weight, given by $s_{\incl} = F/N_{\incl}$,
for the case that only the inclusive sample is used to estimate the trigger rate.
The weight $s_{\incl}$ amounts to $35$~Hz in this example.
As can be seen in Fig.~\ref{fig:weight_trigger_rate}, the addition of samples produced in ranges of $\pThat$ to the inclusive sample reduces the weights.
Lower weights in turn reduce the statistical uncertainties on the trigger rate estimate.
The different maxima in the distribution of stitching weights $s^{I}$ correspond to events 
in which the transverse momentum exchanged between the scattered protons falls into different ranges in $\pThat$.
The spectrum of weights shown in Fig.~\ref{fig:weight_trigger_rate} is plotted before any trigger selection is applied.
As the probability for an event to pass the trigger increases with $\pThat$,
the stitching weights $s^{I}$ are on average smaller for events that pass than for events that fail the trigger selection.
Consequently, the reduction in statistical uncertainties that one obtains by using the exclusive samples and applying the stitching weights 
becomes even more pronounced after the trigger selection is applied (the weight $s_{\incl}$ remains fixed at $35$~Hz).

\begin{figure}
\setlength{\unitlength}{1mm}
\begin{center}
\begin{picture}(180,90)(0,0)
\put(4.5, 4.0){\mbox{\includegraphics*[height=86mm]
  {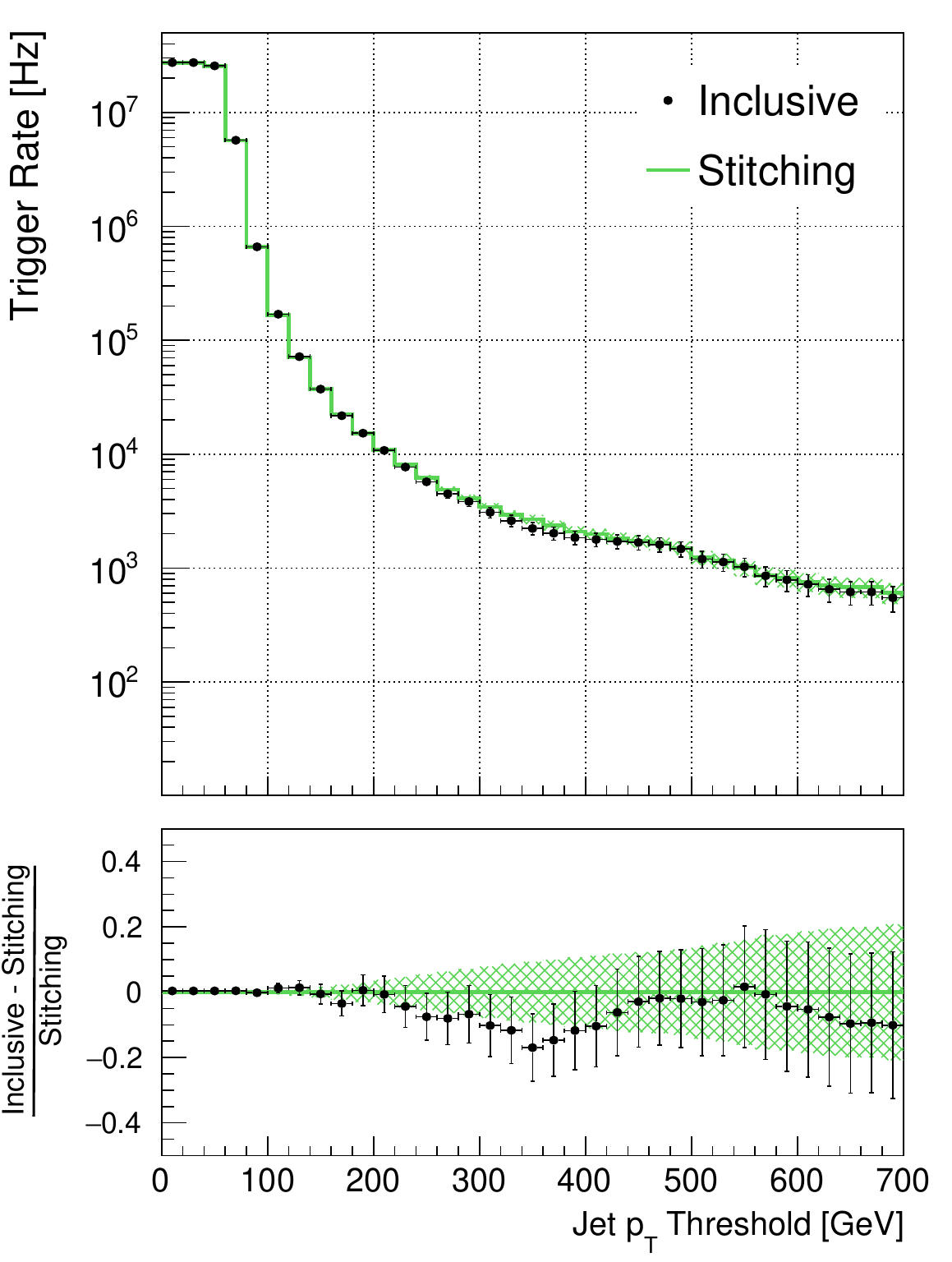}}}
\put(83.5, 4.0){\mbox{\includegraphics*[height=86mm]
  {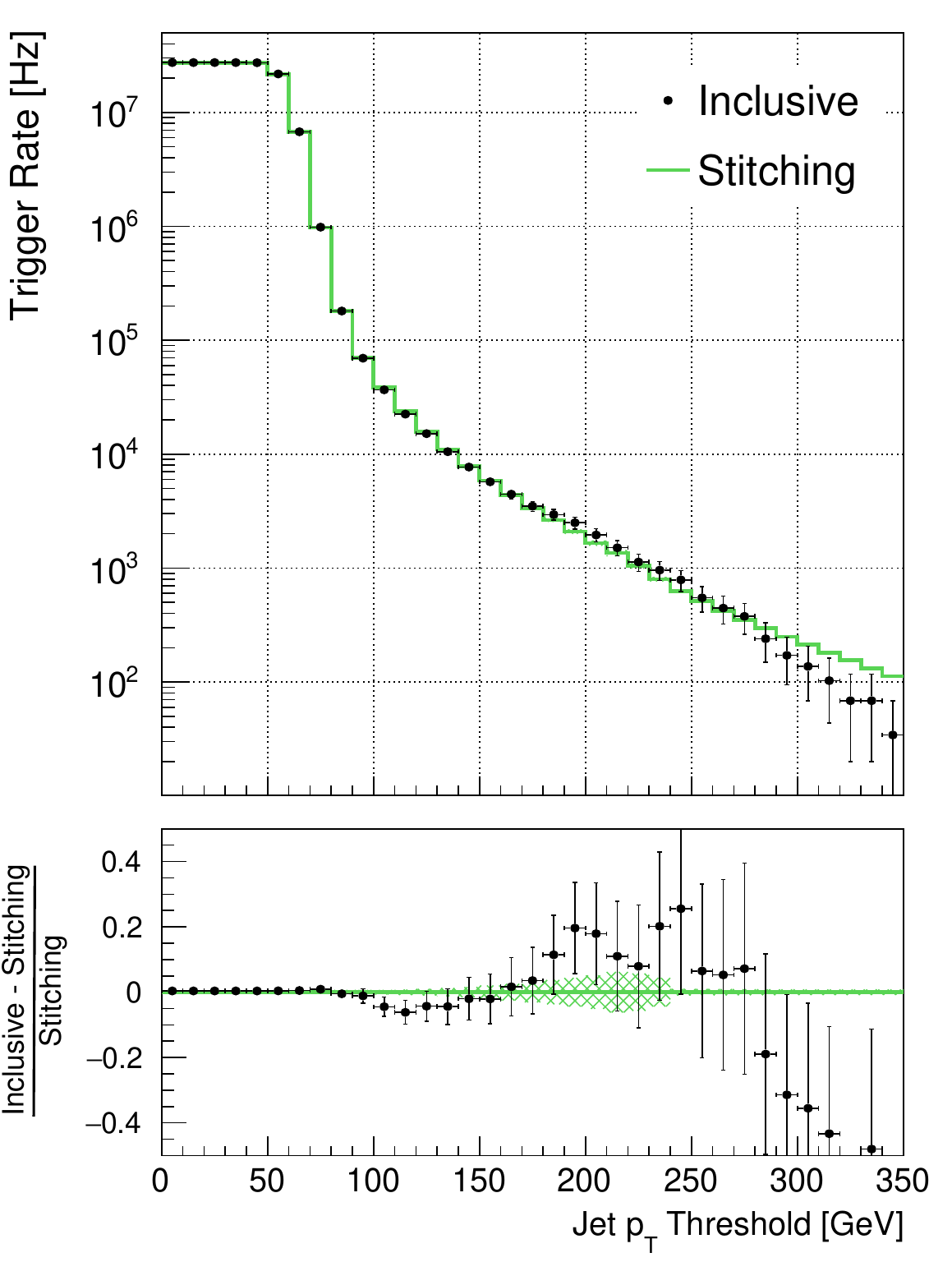}}}
\put(43.0, 0.0){\small (a)}
\put(122.0, 0.0){\small (b)}
\end{picture}
\end{center}
\caption{
  Rate expected for (a) a single jet trigger and (b) a dijet trigger at the HL-LHC, as function of the $\pT$ threshold that is applied to the jets.
}
\label{fig:trigger_rate}
\end{figure}

The rates expected for a single jet trigger and for a dijet trigger at the HL-LHC are shown in Fig.~\ref{fig:trigger_rate}.
The rates are computed as function of the $\pT$ threshold that is applied to the jets. 
In case of the dijet trigger, the same $\pT$ threshold is applied to both jets.
The jets are reconstructed as described in Section~\ref{sec:WJets_vs_Njet}
and are required to be within the geometric acceptance $\vert\eta\vert < 5.0$.
All stable generator-level particles (except neutrinos) originating either from the HS interaction or from any of the PU interactions are used as input to the jet reconstruction.
The statistical uncertainties on the rate estimates obtained from the inclusive sample are represented by error bars,
while those obtained from the sum of inclusive plus exclusive samples are represented by the shaded area.

The rate estimate obtained for the inclusive sample and for the sum of inclusive plus exclusive samples, 
with the stitching weights computed according to Eq.~(\ref{eq:weight_trigger_rate_simplified}),
agree within statistical uncertainties, demonstrating that the estimate of the trigger rate obtained from the stitching procedure is unbiased.
The modest difference between the rate estimates for the dijet trigger with jet $\pT$ thresholds higher than $280$~\GeV
is not statistically significant.
It is important to keep in mind that the trigger rate estimates for adjacent bins are correlated,
because all events that pass the trigger for a given jet $\pT$ threshold also pass the trigger for all lower thresholds.

For both triggers and all jet $\pT$ thresholds,
the statistical uncertainties obtained with the stitching procedure are smaller than the uncertainties obtained in case only the inclusive sample is used.
The reduction in the statistical uncertainties is significantly less pronounced for the single jet trigger than for the dijet trigger, however.
For the latter, the stitching procedure reduces the statistical uncertainties in particular for jet $\pT$ thresholds higher than $100$~\GeV.
The reduction in the statistical uncertainties for the single jet trigger is limited by events with low $\pThat$ that contain a single jet of high $\pT$.
The stitching weights $s^{I}$ for these low $\pThat$ events are not much smaller than the weights for the inclusive sample.
These low $\pThat$ events also cause a ``flattening'' of the single jet trigger rate for jet $\pT$ thresholds higher than $400$~\GeV.
The requirement of a second high $\pT$ jet removes most of these low $\pThat$ events,
with the effect that the dijet trigger rate decreases more rapidly as function of the jet $\pT$ threshold
and the stitching procedure becomes more effective in reducing the statistical uncertainties for the dijet trigger.

\section{Summary}
\label{sec:summary}

The production of MC samples containing a sufficient number of events to allow for a meaningful comparison with the data is often a challenge in modern HEP experiments,
due to the computing resources required to produce and store such samples.
This is particularly true for experiments at the CERN LHC,
firstly because of the large cross sections of relevant processes (\eg DY, $\PW$+jets, and $\Ptop\APtop$+jets production)
and secondly because of the large luminosity delivered by the LHC.

In this paper we have focused on the case that the MC samples have already been produced
and we have presented a procedure that allows to reduce the statistical uncertainties 
by combining MC samples which overlap in PS.
The procedure is based on applying suitably chosen weights to the simulated events.
We refer to the procedure as ``stitching''.

The formalism for computing the stitching weights is general enough to be applied to a variety of use-cases.
When used in physics analyses, the stitching procedure allows to reduce the statistical uncertainties in particular in the tails of distributions.
Examples that document the typical use of the stitching procedure in physics analyses performed by the CMS experiment during LHC Runs $1$ and $2$ have been presented.
The formalism has been extended to the case of estimating trigger rates at the HL-LHC.
Up to  $200$ simultaneous $\Pp\Pp$ collisions are expected per crossing of the proton beams at the HL-LHC.
The distinguishing feature of this application of the stitching procedure is that the same physics process, 
inelastic $\Pp\Pp$ scattering interactions in which a transverse momentum $\pThat$ is exchanged between the protons,
may occur in the ``hard-scatter'' (HS) interaction and in ``pileup'' (PU) interactions.
Our formalism for computing the stitching weight treats the HS and PU interactions on equal footing.

The examples demonstrate that the stitching procedure provides unbiased estimates of event yields and rates as well as of the shapes of distributions.
The reduction in the statistical uncertainties achieved by the stitching method depends on the number of events contained in the MC samples that are subject to the stitching procedure
and ranges from moderate to significant.

\section{Acknowledgements}

This work was supported by the Estonian Research Council grant PRG445.

\appendix

\begin{sidewaystable}
\centering              
\resizebox{0.9\textwidth}{!}{
\def\arraystretch{1.3}
\begin{tabular}{lccccccccccccccc}
Sample                           &  $P^{0}$ &  $P^{1}$ &  $P^{2}$ &  $P^{3}$ &  $P^{4}$ &  $P^{5}$ &  $P^{6}$ &  $P^{7}$ &  $P^{8}$ &  $P^{9}$ &  $P^{10}$         &  $P^{11}$            &  $P^{12}$            & $P^{13}$ & $P^{14}$ \\
\hline
Inclusive                        &  $0.760$ &  $-$     &  $-$     &  $-$     &  $-$     &  $-$     &  $-$     &  $-$     &  $-$     &  $0.156$ &  $6.04\times10^{-3}$ &  $2.51\times10^{-3}$ &  $1.50\times10^{-4}$ &  $4.48\times10^{-6}$ &  $-$ \\
$N_{\jet} = 1$                   &  $-$ &  $-$ &  $-$ &  $-$ &  $-$ &  $-$ &  $-$ &  $-$ &  $-$ &  $0.948$ &  $3.67\times10^{-2}$ &  $1.52\times10^{-2}$ &  $9.08\times10^{-4}$ &  $2.72\times10^{-5}$ &  $1.71\times10^{-6}$ \\
$N_{\jet} = 2$                   &  $-$ &  $-$ &  $-$ &  $-$ &  $-$ &  $-$ &  $-$ &  $-$ &  $-$ &  $-$ &  $-$ &  $-$ &  $-$ &  $-$ &  $-$ \\
$N_{\jet} = 3$                   &  $-$ &  $-$ &  $-$ &  $-$ &  $-$ &  $-$ &  $-$ &  $-$ &  $-$ &  $-$ &  $-$ &  $-$ &  $-$ &  $-$ &  $-$ \\
$N_{\jet} = 4$                   &  $-$ &  $-$ &  $-$ &  $-$ &  $-$ &  $-$ &  $-$ &  $-$ &  $-$ &  $-$ &  $-$ &  $-$ &  $-$ &  $-$ &  $-$ \\
$  70 \leqslant \HT <  100$~\GeV &  $-$ &  $-$ &  $-$ &  $-$ &  $-$ &  $-$ &  $-$ &  $-$ &  $-$ &  $-$ &  $0.261$ &  $-$ &  $-$ &  $-$ &  $-$ \\
$ 100 \leqslant \HT <  200$~\GeV &  $-$ &  $-$ &  $-$ &  $-$ &  $-$ &  $-$ &  $-$ &  $-$ &  $-$ &  $-$ &  $-$ &  $0.109$ &  $-$ &  $-$ &  $-$ \\
$ 200 \leqslant \HT <  400$~\GeV &  $-$ &  $-$ &  $-$ &  $-$ &  $-$ &  $-$ &  $-$ &  $-$ &  $-$ &  $-$ &  $-$ &  $-$ &  $2.43\times10^{-2}$ &  $-$ &  $-$ \\
$ 400 \leqslant \HT <  600$~\GeV &  $-$ &  $-$ &  $-$ &  $-$ &  $-$ &  $-$ &  $-$ &  $-$ &  $-$ &  $-$ &  $-$ &  $-$ &  $-$ &  $5.37\times10^{-3}$ &  $-$ \\
$ 600 \leqslant \HT <  800$~\GeV &  $-$ &  $-$ &  $-$ &  $-$ &  $-$ &  $-$ &  $-$ &  $-$ &  $-$ &  $-$ &  $-$ &  $-$ &  $-$ &  $-$ &  $1.37\times10^{-3}$ \\
$ 800 \leqslant \HT < 1200$~\GeV &  $-$ &  $-$ &  $-$ &  $-$ &  $-$ &  $-$ &  $-$ &  $-$ &  $-$ &  $-$ &  $-$ &  $-$ &  $-$ &  $-$ &  $-$ \\
$1200 \leqslant \HT < 2500$~\GeV &  $-$ &  $-$ &  $-$ &  $-$ &  $-$ &  $-$ &  $-$ &  $-$ &  $-$ &  $-$ &  $-$ &  $-$ &  $-$ &  $-$ &  $-$ \\
$       \HT \geqslant 2500$~\GeV &  $-$ &  $-$ &  $-$ &  $-$ &  $-$ &  $-$ &  $-$ &  $-$ &  $-$ &  $-$ &  $-$ &  $-$ &  $-$ &  $-$ &  $-$ \\

\multicolumn{16}{c}{} \\

Sample                           & $P^{15}$ & $P^{16}$ & $P^{17}$ & $P^{18}$ & $P^{19}$ & $P^{20}$ & $P^{21}$ & $P^{22}$ & $P^{23}$ & $P^{24}$ & $P^{25}$ & $P^{26}$ & $P^{27}$ & $P^{28}$ & $P^{29}$ \\
\hline
Inclusive                        & $9.20\times10^{-8}$ &  $-$ &  $-$ &  $2.84\times10^{-2}$ &  $1.29\times10^{-2}$ &  $9.68\times10^{-3}$ &  $1.23\times10^{-3}$ &  $7.58\times10^{-5}$ &  $1.22\times10^{-5}$ &  $3.49\times10^{-6}$ &  $6.61\times10^{-7}$ &  $-$ &  $1.36\times10^{-3}$ &  $4.07\times10^{-3}$ &  $7.61\times10^{-3}$ \\
$N_{\jet} = 1$                   &  $5.40\times10^{-7}$ &  $-$ &  $-$ &  $-$ &  $-$ &  $-$ &  $-$ &  $-$ &  $-$ &  $-$ &  $-$ &  $-$ &  $-$ &  $-$ &  $-$ \\
$N_{\jet} = 2$                   &  $-$ &  $-$ &  $-$ &  $0.543$ &  $0.247$ &  $0.185$ &  $2.36\times10^{-2}$ &  $1.45\times10^{-3}$ &  $2.33\times10^{-4}$ &  $6.68\times10^{-5}$ &  $1.26\times10^{-5}$ &  $-$ &  $-$ &  $-$ &  $-$ \\
$N_{\jet} = 3$                   &  $-$ &  $-$ &  $-$ &  $-$ &  $-$ &  $-$ &  $-$ &  $-$ &  $-$ &  $-$ &  $-$ &  $-$ &  $8.93\times10^{-2}$ &  $0.267$ &  $0.499$ \\
$N_{\jet} = 4$                   &  $-$ &  $-$ &  $-$ &  $-$ &  $-$ &  $-$ &  $-$ &  $-$ &  $-$ &  $-$ &  $-$ &  $-$ &  $-$ &  $-$ &  $-$ \\
$  70 \leqslant \HT <  100$~\GeV &  $-$ &  $-$ &  $-$ &  $-$ &  $0.557$ &  $-$ &  $-$ &  $-$ &  $-$ &  $-$ &  $-$ &  $-$ &  $-$ &  $0.176$ &  $-$ \\
$ 100 \leqslant \HT <  200$~\GeV &  $-$ &  $-$ &  $-$ &  $-$ &  $-$ &  $0.422$ &  $-$ &  $-$ &  $-$ &  $-$ &  $-$ &  $-$ &  $-$ &  $-$ &  $0.331$ \\
$ 200 \leqslant \HT <  400$~\GeV &  $-$ &  $-$ &  $-$ &  $-$ &  $-$ &  $-$ &  $0.200$ &  $-$ &  $-$ &  $-$ &  $-$ &  $-$ &  $-$ &  $-$ &  $-$ \\
$ 400 \leqslant \HT <  600$~\GeV &  $-$ &  $-$ &  $-$ &  $-$ &  $-$ &  $-$ &  $-$ &  $9.09\times10^{-2}$ &  $-$ &  $-$ &  $-$ &  $-$ &  $-$ &  $-$ &  $-$ \\
$ 600 \leqslant \HT <  800$~\GeV &  $-$ &  $-$ &  $-$ &  $-$ &  $-$ &  $-$ &  $-$ &  $-$ &  $5.94\times10^{-2}$ &  $-$ &  $-$ &  $-$ &  $-$ &  $-$ &  $-$ \\
$ 800 \leqslant \HT < 1200$~\GeV &  $1.06\times10^{-4}$ &  $-$ &  $-$ &  $-$ &  $-$ &  $-$ &  $-$ &  $-$ &  $-$ &  $4.02\times10^{-2}$ &  $-$ &  $-$ &  $-$ &  $-$ &  $-$ \\
$1200 \leqslant \HT < 2500$~\GeV &  $-$ &  $3.07\times10^{-4}$ &  $-$ &  $-$ &  $-$ &  $-$ &  $-$ &  $-$ &  $-$ &  $-$ &  $2.69\times10^{-2}$ &  $-$ &  $-$ &  $-$ &  $-$ \\
$       \HT \geqslant 2500$~\GeV &  $-$ &  $-$ &  $-$ &  $-$ &  $-$ &  $-$ &  $-$ &  $-$ &  $-$ &  $-$ &  $-$ &  $2.15\times10^{-2}$ &  $-$ &  $-$ &  $-$ \\

\multicolumn{16}{c}{} \\

Sample                           & $P^{30}$ & $P^{31}$ & $P^{32}$ & $P^{33}$ & $P^{34}$ & $P^{35}$ & $P^{36}$ & $P^{37}$ & $P^{38}$ & $P^{39}$ & $P^{40}$ & $P^{41}$ & $P^{42}$ & $P^{43}$ & $P^{44}$ \\
\hline
Inclusive                        &  $1.94\times10^{-3}$ &  $1.70\times10^{-4}$ &  $3.06\times10^{-5}$ &  $1.00\times10^{-5}$ &  $1.99\times10^{-6}$ &  $-$ &  $1.39\times10^{-5}$ &  $3.01\times10^{-4}$ &  $3.25\times10^{-3}$ &  $2.83\times10^{-3}$ &  $5.76\times10^{-4}$ &  $1.61\times10^{-4}$ &  $7.27\times10^{-5}$ &  $2.17\times10^{-5}$ &  $-$ \\
$N_{\jet} = 1$                   &  $-$ &  $-$ &  $-$ &  $-$ &  $-$ &  $-$ &  $-$ &  $-$ &  $-$ &  $-$ &  $-$ &  $-$ &  $-$ &  $-$ &  $-$ \\
$N_{\jet} = 2$                   &  $-$ &  $-$ &  $-$ &  $-$ &  $-$ &  $-$ &  $-$ &  $-$ &  $-$ &  $-$ &  $-$ &  $-$ &  $-$ &  $-$ &  $-$ \\
$N_{\jet} = 3$                   &  $0.127$ &  $1.12\times10^{-2}$ &  $2.00\times10^{-3}$ &  $6.57\times10^{-4}$ &  $1.30\times10^{-4}$ &  $-$ &  $-$ &  $-$ &  $-$ &  $-$ &  $-$ &  $-$ &  $-$ &  $-$ &  $-$ \\
$N_{\jet} = 4$                   &  $-$ &  $-$ &  $-$ &  $-$ &  $-$ &  $-$ &  $1.93\times10^{-3}$ &  $4.17\times10^{-2}$ &  $0.452$ &  $0.394$ &  $8.00\times10^{-2}$ &  $2.23\times10^{-2}$ &  $1.01\times10^{-2}$ &  $3.02\times10^{-3}$ &  $1.09\times10^{-4}$ \\
$  70 \leqslant \HT <  100$~\GeV &  $-$ &  $-$ &  $-$ &  $-$ &  $-$ &  $-$ &  $-$ &  $1.30\times10^{-2}$ &  $-$ &  $-$ &  $-$ &  $-$ &  $-$ &  $-$ &  $-$ \\
$ 100 \leqslant \HT <  200$~\GeV &  $-$ &  $-$ &  $-$ &  $-$ &  $-$ &  $-$ &  $-$ &  $-$ &  $0.142$ &  $-$ &  $-$ &  $-$ &  $-$ &  $-$ &  $-$ \\
$ 200 \leqslant \HT <  400$~\GeV &  $0.315$ &  $-$ &  $-$ &  $-$ &  $-$ &  $-$ &  $-$ &  $-$ &  $-$ &  $0.460$ &  $-$ &  $-$ &  $-$ &  $-$ &  $-$ \\
$ 400 \leqslant \HT <  600$~\GeV &  $-$ &  $0.204$ &  $-$ &  $-$ &  $-$ &  $-$ &  $-$ &  $-$ &  $-$ &  $-$ &  $0.691$ &  $-$ &  $-$ &  $-$ &  $-$ \\
$ 600 \leqslant \HT <  800$~\GeV &  $-$ &  $-$ &  $0.149$ &  $-$ &  $-$ &  $-$ &  $-$ &  $-$ &  $-$ &  $-$ &  $-$ &  $0.783$ &  $-$ &  $-$ &  $-$ \\
$ 800 \leqslant \HT < 1200$~\GeV &  $-$ &  $-$ &  $-$ &  $0.116$ &  $-$ &  $-$ &  $-$ &  $-$ &  $-$ &  $-$ &  $-$ &  $-$ &  $0.838$ &  $-$ &  $-$ \\
$1200 \leqslant \HT < 2500$~\GeV &  $-$ &  $-$ &  $-$ &  $-$ &  $8.09\times10^{-2}$ &  $-$ &  $-$ &  $-$ &  $-$ &  $-$ &  $-$ &  $-$ &  $-$ &  $0.884$ &  $-$ \\
$       \HT \geqslant 2500$~\GeV &  $-$ &  $-$ &  $-$ &  $-$ &  $-$ &  $4.77\times10^{-2}$ &  $-$ &  $-$ &  $-$ &  $-$ &  $-$ &  $-$ &  $-$ &  $-$ &  $0.926$ \\
\hline
\end{tabular}
}
\caption{
  Probabilities $P_{\incl}^{i}$ and $P_{j}^{i}$ for the events in the inclusive and exclusive $\PW$+jets samples 
  to populate the different PS regions $i$.
  The definition of the PS regions $i$ in the plane of $N_{\jet}$ versus $\HT$ is shown in Fig.~\ref{fig:regions_WJets_vs_Njet_and_HT}.
  A hyphen ($-$) indicates that PS regions $i$ are not populated by a given sample $j$. The presence of a hyphen is equivalent to the probability $P_{j}^{i}$ being zero.
}
\label{tab:probabilities_WJets_vs_Njet_and_HT}
\end{sidewaystable}

\begin{table}
\centering
\def\arraystretch{1.3}
\begin{tabular}{l|c|c|c|c|c}
                                & $N_{\jet} = 0$ & $N_{\jet} = 1$      & $N_{\jet} = 2$      & $N_{\jet} = 3$      & $N_{\jet} = 4$       \\ 
\hline
               $\HT < 70$~\GeV   & $2870$         & $1420$              &  $718$              &  $359$              &  $176$  \\
  $70 \leqslant \HT < 100$~\GeV  & $-$            &  $714$              &  $478$              &  $287$              &  $157$  \\
 $100 \leqslant \HT < 200$~\GeV  & $-$            &  $472$              &  $357$              &  $238$              &  $141$  \\
 $200 \leqslant \HT < 400$~\GeV  & $-$            &  $283$              &  $237$              &  $178$              &  $118$  \\
 $400 \leqslant \HT < 600$~\GeV  & $-$            &  $158$              &  $144$              &  $120$              & $89.6$  \\
 $600 \leqslant \HT < 800$~\GeV  & $-$            & $83.8$              & $78.6$              & $71.0$              & $60.0$  \\
 $800 \leqslant \HT < 1200$~\GeV & $-$            & $36.0$              & $35.1$              & $33.7$              & $31.2$  \\
$1200 \leqslant \HT < 2500$~\GeV & $-$            & $16.3$              & $15.9$              & $15.4$              & $15.3$  \\
       $\HT \geqslant 2500$~\GeV & $-$            &    $-$              & $11.2$              & $11.2$              & $10.9$  \\
\end{tabular}
\caption{
  Stitching weights $s^{i}$ for the case that the inclusive and exclusive $\PW$+jets samples 
  given in Tables~\ref{tab:samples_and_probabilities_WJets_vs_Njet} and~\ref{tab:samples_WJets_vs_Njet_and_HT}
  are stitched based on the observables $N_{\jet}$ and $\HT$.
  The weights are computed for an integrated luminosity of $140\fbinv$.
  Events with $N_{\jet} = 0$ all have $\HT < 70$~\GeV,
  and, similarly, events with $N_{\jet} = 1$ all have $\HT < 2500$~\GeV.
  Hence, the stitching weights cannot be computed for the eight PS regions with $N_{\jet} = 0$ and $\HT \geqslant 70$~\GeV 
  and for the single PS region with $N_{\jet} = 1$ and $\HT \geqslant 2500$~\GeV.
  These stitching weights are not needed, since there are no events in these regions.
  The corresponding PS regions are indicated by a hyphen ($-$) in the table.
}
\label{tab:weights_WJets_vs_Njet_and_HT}
\end{table}

\section{Appendix}
\label{sec:appendix}

In this section, we detail the application of the least-squares method~\cite{Cowan:1998ji}
to the computation of the probabilities $P_{\incl}^{i}$ and $P_{j}^{i}$
in the example given in Section~\ref{sec:WJets_vs_Njet_and_HT}.
The aim is to make the computation of these probabilities less prone to statistical fluctuations in the MC samples.

Our use of the least-squares method is based on assuming the following relation to hold:
\begin{equation}
\sigma_{\incl} \, P_{\incl}^{i} = \sigma_{j} \, P_{j}^{i} \quad \forall \, j \, ,
\label{eq:lambda1}
\end{equation}
except for statistical fluctuations on the $P_{\incl}^{i}$ and $P_{j}^{i}$.
The $P_{\incl}^{i}$ and $P_{j}^{i}$ are obtained by determining the fraction of events in the inclusive and exclusive MC samples that fall into PS region $i$.
The symbol $j$ in Eq.~(\ref{eq:lambda1}) refers to those $N_{\jet}$- and $\HT$-samples that cover the PS region $i$.
The symbol $\sigma_{j}$ refers to the fiducial cross section corresponding to the sample $j$.
We denote the unknown true value of the left-hand side (and equivalently of the right-hand side) of Eq.~(\ref{eq:lambda1}) by the symbol $\lambda_{i}$
and use the symbols $r_{\incl}^{i}$ and $r_{j}^{i}$ to refer to the deviations (``residuals''), caused by statistical fluctuations,
between the true values of the probabilities $P_{\incl}^{i}$ and $P_{j}^{i}$ and the values obtained using the MC samples.
We further use the symbols $e_{\incl}^{i}$ and $e_{j}^{i}$ to denote the expected statistical fluctuations (``errors'') of these probabilities.
According to the least-squares method,
the best estimate for the value of $\lambda_{i}$ is given by the solution to the following system of equations:
\begin{eqnarray*}
\sigma_{\incl} \, \left( P_{\incl}^{i} + r_{\incl}^{i} \right) - \lambda_{i} & = & 0 \quad \mbox{and} \\
\sigma_{j} \, \left( P_{j}^{i} + r_{j}^{i} \right) - \lambda_{i} & = & 0 \quad \forall \, j \, ,
\end{eqnarray*}
subject to the condition that the sum of residuals:
\begin{equation*}
\left( \frac{r_{\incl}^{i}}{e_{\incl}^{i}} \right)^{2} \, + \, \sum_{j} \left( \frac{r_{j}^{i}}{e_{j}^{i}} \right)^{2}
\end{equation*}
attains its minimal value.
The expected statistical fluctuations $e_{\incl}^{i}$ and $e_{j}^{i}$ of the probabilities $P_{\incl}^{i}$ and $P_{j}^{i}$
are given by the standard errors of the Binomial distribution~\cite{Cowan:1998ji}:
\begin{equation*}
e_{\incl}^{i} = \sqrt{\frac{P_{\incl}^{i} \, (1 - P_{\incl}^{i})}{N_{\incl}}} \qquad \mbox{and} \qquad e_{j}^{i} = \sqrt{\frac{P_{j}^{i} \, (1 - P_{j}^{i})}{N_{j}}} \, .
\end{equation*}
The fluctuations decrease proportional to the inverse of the square-root of the number of events in the MC samples.
The solution for $\lambda_{i}$ is given by the expression:
\begin{equation}
\lambda_{i} = \frac{\alpha_{\incl}^{i} \, \sigma_{\incl} \, P_{\incl}^{i} + \sum_{j} \alpha_{j}^{i} \, \sigma_{j} \, P_{j}^{i}}{\alpha_{\incl}^{i} + \sum_{j} \alpha_{j}^{i}} \, ,
\label{eq:lambda2}
\end{equation}
from which the probabilities $P_{\incl}^{i} = \lambda_{i}/\sigma_{\incl}$ and $P_{j}^{i} = \lambda_{i}/\sigma_{j}$ follow.
The symbols $\alpha_{\incl}^{i}$ and $\alpha_{j}^{i}$ are defined as:
\begin{equation*}
\alpha_{\incl}^{i} = \frac{1}{\left( \sigma_{\incl} \, e_{\incl}^{i} \right)^{2}} \qquad \mbox{and} \qquad \alpha_{j}^{i} = \frac{1}{\left( \sigma_{j} \, e_{j}^{i} \right)^{2}}
\end{equation*}
and act as ``weights'' in the expression on the right-hand side of Eq.~(\ref{eq:lambda2}),
which has the form of a weighted average.
We use the symbols $\alpha_{\incl}^{i}$ and $\alpha_{j}^{i}$ to refer to these weights,
in order to distinguish them from the weights $w_{\incl}^{k}$ and $w_{j}^{k}$ computed by the MC generator program
and from the stitching weights $s^{i}$.

\clearpage

\bibliography{mcStitching}
%\endgroup

\end{document}